\documentclass[12pt,a4paper]{article}
\pdfoutput=1
\usepackage[utf8]{inputenc}
\usepackage[T1]{fontenc}
\usepackage{amsmath}
\usepackage{amsfonts}
\usepackage{amssymb}
\usepackage{fullpage}
\usepackage{bbold}
\usepackage{xcolor}
\usepackage{lscape}
\usepackage{tabularx}
\usepackage{url}
\usepackage{natbib}
\usepackage[font={small}]{caption}
\usepackage{mathtools}
\usepackage[affil-it]{authblk}
\usepackage{longtable}
\usepackage{float}
\usepackage[hmargin=1in,vmargin={1in,1in}]{geometry}
\usepackage[space]{grffile}
\usepackage{caption}
\usepackage{subcaption}
\usepackage{graphicx}

\title{Wright meets Markowitz: \\
How standard portfolio theory changes when assets are technologies following experience curves}

\author[a, b]{Rupert Way \thanks{rupert.way@smithschool.ox.ac.uk}}
\author[a, b, c]{Fran\c{c}ois Lafond \thanks{francois.lafond@inet.ox.ac.uk}}
\author[d, e]{Fabrizio Lillo \thanks{fabrizio.lillo@sns.it}}
\author[f]{Valentyn Panchenko \thanks{v.panchenko@unsw.edu.au}}
\author[a, g, h, i]{J. Doyne Farmer \thanks{doyne.farmer@inet.ox.ac.uk}}
\affil[a]{Institute for New Economic Thinking at the Oxford Martin School, University of Oxford, Oxford OX2 6ED, UK}
\affil[b]{Smith School of Enterprise and the Environment, University of Oxford, Oxford, OX1 3QY}
\affil[c]{Oxford Martin School Programme on Technological and Economic Change, University of Oxford, OX1 3BD}
\affil[d]{Scuola Normale Superiore, Piazza dei Cavalieri 7, 56126 Pisa, Italy}
\affil[e]{CADS, Center for Analysis, Decisions, and Society, Human Technopole, Milano, 20156, Italy}
\affil[f]{School of Economics, UNSW Business School, Sydney NSW 2052, Australia}
\affil[g]{Mathematical Institute, University of Oxford, Oxford OX1 3LP, UK}
\affil[h]{Department of Computer Science, University of Oxford, Oxford OX1 3QD, UK}
\affil[i]{Santa-Fe Institute, Santa Fe, NM 87501, USA}

\begin{document}

\maketitle

\newpage

\begin{abstract}
We consider how to optimally allocate investments in a portfolio of competing technologies using the standard mean-variance framework of portfolio theory.
We assume that technologies follow the empirically observed relationship known as Wright's law, also called a ``learning curve'' or ``experience curve'', which postulates that costs drop as cumulative production increases.
This introduces a positive feedback between cost and investment that complicates the portfolio problem, leading to multiple local optima, and causing a trade-off between concentrating investments in one project to spur rapid progress vs.\ diversifying over many projects to hedge against failure.
We study the two-technology case and characterize the optimal diversification in terms of progress rates, variability, initial costs, initial experience, risk aversion, discount rate and total demand.
The efficient frontier framework is used to visualize technology portfolios and show how feedback results in nonlinear distortions of the feasible set.
For the two-period case, in which learning and uncertainty interact with discounting, we compare different scenarios and find that the discount rate plays a critical role.
\end{abstract}

\newpage

\section{Introduction}

There is a fundamental trade-off, encountered throughout life, between investing enough effort in any one activity to make rapid progress, and diversifying effort over many projects simultaneously to hedge against failure.
On the one hand, by focusing on a single task one can quickly accumulate experience, become an expert, and reap rewards more efficiently.
But on the other, unforeseen circumstances can impede progress or make the rewards less valuable, so it may be wise to maintain progress on several fronts at once, even if individually slower.
This brings to mind the familiar adage ``don't put all your eggs in one basket'', and at first glance appears very similar to the question of how diversified a portfolio of financial investments should be.
However there is a key difference between the financial portfolio setting and the type of problem considered in this paper, which is that here \textit{learning} is involved: the more effort we invest in one area, the more \textit{effective} that effort becomes --- so we may want to put all our eggs in one basket after all.

The dilemma is ubiquitous, and is understood intuitively by us all as we learn new skills, engage in new projects, and attempt to plan for the future.
For example, consider trying to decide how many courses to take in university; or how many languages, musical instruments, sports, or web application frameworks to learn.
Focusing on one, or just a few, allows us to gain expertise and reach a more rewarding phase of activity sooner.
Or at the organisational level, firms and governments must decide how many, and which, strategic and technological capabilities to develop.
We present a simple model for understanding this trade-off, and show how it is related to the optimal diversification problem for financial assets.

The reason this decision framework is of particular interest is that, despite its simplicity, it shares several important features with the question of how to allocate investments among competing technologies.
This is because often, in the long run, scientific advances and knowledge gains mean that performance-weighted technology investment costs decrease as cumulative deployment increases.
Put simply, in such cases the more we invest in a technology (whether at the R\&D, deployment or any other stage) the more effective the technology becomes at delivering the same output, so future investment costs are lower, per unit of output\footnote{Though note that many technologies do not exhibit such decreasing costs at all, and also that the effect (when it exists) is far less evident in mature technologies since so much experience has already accrued.}.
Hence, in order to achieve certain long term technological goals, understanding the correct allocation of investments among available substitute technologies is vital.
The specific question we have in the back of our minds is how to allocate funding over potential clean energy technologies to accelerate the transition to a net zero carbon economy --- should we invest in solar photovoltaics, or offshore wind, or next generation nuclear, or carbon capture and storage, or a little bit in each?

Despite this high-level motivation, here we focus in on a very simple conceptual model representing the underlying trade-off.
The key assumption we make is that increased cumulative investment in a technology \textit{leads to} reduced investment costs (but with some degree of uncertainty).
In reality this causal mechanism is not so straightforward, and there are many other complicating factors, such as correlations between projects and spillovers (incoming and outgoing) of various kinds.
However, while stating the caveats clearly, we set aside these issues for now and just focus on the core problem, which is to find the optimal risk-averse investment in competing technologies following experience curves. This setting brings together the specialization incentives of the learning curve model with the diversification incentives of modern portfolio theory, allowing us to characterize the optimal solution to the trade-off between diversifying and specializing.

Our approach is to consider multiple independent technologies (two), increasing returns to investment (through experience curves), uncertainty (cost is a stochastic process), and a risk-averse decision maker (who minimizes a mean-variance value function). Since the two technologies follow stochastic processes, diversification tends to reduce the risks, but at the same time increasing returns tend to favour specialization.
Investing in one option drives down its marginal cost, making it more and more attractive, but \emph{ex-ante} uncertainty in future benefits from learning suggests that diversifying can limit the risk of over-investing in a technology that eventually shows a poor performance. We characterize optimal investment as a function of learning characteristics (rate and uncertainty of learning), initial conditions (cost competitiveness and accumulated experience), risk aversion, discount rate and the level of demand.
We focus primarily on the one-period investment decision, but also consider the extension to two periods.

In classical (Markowitz) portfolio theory, the optimal allocation of investments is unique. In general, the further a portfolio is from the optimum, the worse is its value. In contrast, when the positive feedback of endogenous technological progress is strong, it is better to invest mostly in \emph{either} of the two options than to split investment more evenly. Except in some knife-edge cases, one of the two specialized portfolios is better than the other, but which one is best depends on the parameters. As a result, a small change in one of the parameters can result in the optimal portfolio being completely different.

In general we characterize three different regimes. In the first regime, one technology is so much better than the other that it dominates the portfolio entirely.
This happens either because there is no risk aversion (so we revert to the classic deterministic learning curve winner-takes-all scenario) or because the relative advantage of one technology in terms of initial conditions or speed and uncertainty of learning is very strong.
In the second regime, the optimal portfolio features unambiguous diversification, that is, the objective function has a unique optimum corresponding to a balanced mix of technologies.
In the third regime, there are two local optima of similar value, corresponding to quite different investment policies. As parameters change, the transition of the global optimum from one local optimum to the other is abrupt. In other words, in this critical region, a small change in a parameter causes a large change in the optimal policy.  

We show what this finding implies for the theory of path-dependence and lock-in, and characterize lock-in as a situation where investing in a fast-learning technology is not currently optimal, but it \textit{would be} if a higher level of demand existed. Intuitively, whether or not one should attempt to bring a technology down its learning curve depends on the size of the market. For some parameter values though, the transition is sharp \--- there exists a critical level of demand below which investment in the fast-learning technology is limited, and above which it becomes dominant (i.e.\ the global optimum switches from one local optimum to the other).

We show analytically that a Markowitz-like case may be recovered in two different ways.
First, when there is no learning increasing returns are absent and it becomes highly unlikely that one would want to specialize entirely. (Specialization is still possible, but only because one technology is currently much better than the other, \textit{not} because investing in it makes it better.)
Second, when future demand is very small, compared to the current level, the potential for learning is insignificant and therefore investment is never enough for increasing returns to really matter.

Our results relate to several different branches of literature. 
First of all we are motivated by the optimal energy systems, energy transition and climate change literatures. It is now clear that to avoid a rise in temperatures that would have ``dangerous'' effects we must attain net zero carbon emissions, and one of the main factors involved in this is a switch to a clean energy system. However, dirty energy technologies are currently considered to be both cheap and convenient (in some ways due to legacy energy system design and infrastructure, and societal embeddedness), whereas alternatives are still expensive, even though their cost is falling, sometimes very fast. In this context the questions arise of what costs will be in the future, which decisions affect these costs, and what is the best investment or tax/subsidy policy. Energy systems are highly complex\footnote{Each energy source has specific infrastructure building time (e.g.\ nuclear takes a long time), can be intermittent or not (e.g.\ solar energy is not produced at night), has specific transport, storage and safety conditions, etc.} and energy experts generally rely on highly detailed models of the energy production and consumption mix. To include endogenous technological change in these models, a simple solution which has been widely adopted (and criticized) is that of experience curves \citep{gritsevskyi2000modeling,barreto2004endogenizing,alberth2007climate,criqui2015mitigation,webster2015should}. However, these models end up being very complex so that optimal policies are very hard to determine and understand. Numerical methods have to be used, and it is not always the case that the global optimum is found. Analytical approaches for these complex models generally have to assume a deterministic setting, so that the increasing returns induced by the learning curve lead to full specialization, see for instance \citet{wagner2014optimal}.
In this paper we only wish to understand the fundamental trade-off involved in technology investment: diversification against specialization, and the risk of lock-in in systems with path dependent, self-reinforcing dynamics.
Therefore, we do not attempt to provide a realistic model of the energy system or a direct empirical application of our results, and instead focus on a theoretical contribution at the intersection of the learning-by-doing and portfolio literatures.

To model technological progress, we use a very specific parametric model. Technological progress is not perfectly predictable, but in many detailed empirical cases it has been found that unit costs tend to decrease by a constant percentage every time cumulative production doubles. Subject to some uncertainty about future shocks, the cost of a technology follows an experience curve which is technology-specific. This relationship between unit cost and cumulative investment has been observed for a long time \citep{wright1936factors, alchian1963reliability,thompson2012relationship} and is generally explained by the fact that during production learning-by-doing takes place\footnote{While learning-by-doing often refers to labour force or organisational learning in particular, two other related and noteworthy sources of increasing returns include economies of scale, which depend only on current (not cumulative) production levels, and network externalities, which depend on the number of consumers or other producers joining or using the same network or technology. The key feature of experience curves, however, is that performance increases depend on the growth of total experience (cumulative production), not on the growth of production. We think of experience curves as capturing all experience-related effects, including, but not limited to learning-by-doing.}. Starting with \citet{arrow1962economic}, a large literature has developed to analyse the consequences of this relationship for pricing and output decisions \citep{rosen1972learning,spence1981learning,mazzola1997stochastic}. Learning-by-doing decreases marginal cost, which gives an advantage to size and may encourage predatory pricing \citep{cabral1994learning} or legitimize the protection of infant industry from international competition \citep{dasgupta1988learning}. When one considers a single firm operating a single technology, learning-by-doing generates irreversibilities and creates an incentive to delay investment. The optimal investment dynamics can be characterized using the theory of real options \citep{brueckner1983optimal,majd1989learning,della2012optimal}. In general, the literature does not study multiple technologies at the same time; and when it does, for instance when characterizing the social optimum for a multi-firm sector, it is generally in the absence of uncertainty. Given our motivation to understand optimal investment in energy technologies, which are very diverse and uncertain, we turn to another branch of literature which has dealt in detail with investment in multiple uncertain assets, that is modern portfolio theory \citep{markowitz1952portfolio}.

Modern portfolio theory considers a risk-averse decision maker who wishes to invest in financial assets. The key result of portfolio theory is that there exists an optimal way of combining assets in a portfolio such that expected returns are maximized, conditional on a given level of risk (or that risk is minimized, conditional on a given level of expected returns). We argue that this idea is well suited for thinking about technology investment, and we borrow from portfolio theory the mean-variance value function (in our case, both expected costs and variance of the portfolio have to be minimized). For simplicity, however,
we generally assume that technologies are uncorrelated. As opposed to a ``learning curve technology'', a key property of a financial asset is that investing in it does not change its value, although there are some important exceptions\footnote{One is the situation in which market impact is considered. Market impact acknowledges that trading large quantities simply violates the atomicity assumption, so that one's choice of quantities demanded or supplied affects the price. In this case, this is a \emph{negative} feedback and the literature has focused on finding optimal liquidation strategies \citep{almgren2001optimal,he2005dynamic}. Another situation in which financial portfolios incorporate feedback effects is when learning about an asset is taken into account. An investor who is familiar with a particular asset makes more precise estimates of expected returns, so that this asset is relatively more valuable than other assets \citep{boyle2012keynes}. In turn, holding a lot of a particular asset makes information acquisition about that asset more valuable, which can generate a positive feedback that encourages specialization \citep{van2010information}.}. We recover a classical portfolio setup when the learning parameter is zero or when the total market size is very small compared to the initial production.

Besides our general motivation (energy systems) and the two major ingredients of our model (experience curves and mean-variance portfolio theory), our setup relates to a large literature dealing with optimal control of stochastic processes, which goes well beyond economics and operation research. Of more direct interest are the applications to technology, R\&D and innovation problems, where the questions of increasing returns and lock-in are more salient. 
When investing in an option makes it better and better, history matters. \citet{atkinson1969new} already pointed out that localized technological progress, an important source of which is learning-by-doing, would justify investing in a technology that is not yet the cheapest. In the technology choice literature, it is well known that increasing returns and uncertainty may result in situations where poor technological options dominate \citep{david1985clio}. In a model of two competing standards operating under network externalities, \cite{arthur1989competing} showed that if chance favors an intrinsically worse option early on, this option's accumulated experience gives it an edge for obtaining the marginal consumer. As this advantage accumulates, it may forever exceed the benefits from switching to the intrinsically better option. In this context a policy maker is interested in a policy that optimally explores the merit of different options before making a final choice. \citet{cowan1991tortoises} characterized a social planner's optimal decision in a two arm bandit framework, where there is a choice between one of two technologies at every period. In this model, there exists an optimal policy known as the Gittins index, but according to this policy eventually a single technology will be chosen. Thus early bad luck may induce the social planner to lock in the wrong technology. While this theoretical literature often refers to learning-by-doing\footnote{When increasing returns are from the consumer side, typically as in \citet{arthur1989competing}, they are generally motivated as learning-by-using following \citet{rosenberg1982inside}.}, it attempts to model other forms of increasing returns all at once and therefore does not model more explicitly how cost decreases with investment. \citet{zeppini2015discrete} considered learning curves for clean and dirty technologies in a discrete choice framework, with social interactions as an additional source of increasing returns to adoption and lock-in. He found that policies inducing the clean technology to progress down its learning curve faster have greater potential to induce smooth technological transitions, as opposed to traditional policies such as a pollution tax which can work only by being large enough to induce an equilibrium shift. Finally, another branch of literature has contrasted the benefits of increasing returns against the benefits of technological diversity by assuming that further technological progress takes place through recombination. This implies that there is some value in giving up on increasing returns from specialization and keeping a range of diverse technologies available for further re-combination \citep{van2008optimal,zeppini2013optimal}.

The paper is organized as follows. Section \ref{section:model} defines the stochastic process for the experience curves and the optimization problem in the one-period case, and shows how it relates to Markowitz portfolios.
Section \ref{section:t1} presents the main results of the optimization and shows under which conditions diversification is optimal. It also analyzes in detail the objective function by characterizing how the number and nature of optima changes with underlying parameter values, and studies the effect of total demand.
Section \ref{section:efficient_frontier} returns to the comparison of financial and technology portfolios and shows how the efficient frontier changes when technologies are introduced.
Section \ref{section:lockin} establishes conditions to escape lock-in by studying the case where a mature, cheap but slow-learning technology dominates the market but faces competition from a young, expensive but fast-learning challenger.
Section \ref{section:two_period_model} introduces the multi-period model and explores how discounting interacts with risk aversion and learning in a two-period setting.
Finally, Section \ref{section:conclusion} concludes.

\section{One-period model}
\label{section:model}

Consider the development of a single technology over one time period. The unit cost of the technology at time $t$ is $c_t$ (measured in $ \$ /unit$), and its cumulative production\footnote{We use the terms investment and production interchangeably throughout the one-period model presentation. Generally the literature considers production, although the original paper by \citet{arrow1962economic} used investment. Here we are looking only one step ahead so this is not an important difference.} (measured in $units$) is $z_t$.
Let $t=0$ be the present time and $t=1$ be some given future time. The current unit cost is $c_0$ and current cumulative production is $z_0$. Production during the period is $q$, and the cumulative production at $t=1$ is $z_1 = z_0 + q$. We first present the stochastic model for a single technology then consider a portfolio of two such technologies.

\subsection{Wright's law}

The standard form of the experience curve is
\begin{equation}
c_t \propto {z_t}^{-\alpha},
\label{eq:Wrightlawdet}
\end{equation}
where the constant $\alpha$ is the experience exponent (or Wright exponent) for this technology. This leads to two related concepts often used in the literature: the ``progress ratio'' is defined as the relative cost level seen after each doubling of cumulative production, $PR = 2^{-\alpha}$, while the ``learning rate'' is defined as the relative cost reduction seen after each such doubling, $LR=1-2^{-\alpha}$.
\citet{dutton1984treating} report learning rates from different studies and find that the vast majority lie between 5\% and 40\%, corresponding to values of $\alpha$ lying approximately within the range $(0.07,0.7)$. However, commodities such as minerals and fossil fuels mostly have $\alpha \approx 0$ since they do not exhibit a significant cost decrease over the long run \citep{newbold2005well,mcnerney2011historical}. The power law relationship between cost and cumulative production was first noted by \citet{wright1936factors} in the context of the production of airplanes, so we call it Wright's law. Since then it has been found to describe the available evidence for a number of technologies fairly well \citep{nagy2013statistical}. In contrast to a large part of the theoretical literature on experience curves, which deals only with the deterministic form, we model uncertainty explicitly. To do this we make the future cost stochastic by assuming additive noise $\eta$ on the log-first-difference version of Eq.\ (\ref{eq:Wrightlawdet}):
\begin{equation}
\label{log_first_diff_eqn}
\log(c_{1})- \log (c_0) = -\alpha \big[ \log(z_{1}) - \log (z_0) \big] +\eta.
\end{equation}
This equation models a situation where, over the course of one period, an underlying linear trend in log-log space advances according to Wright's law, but then is hit by a random shock. It is one of the simplest possible ways of incorporating uncertainty in the experience curve model, chosen here specifically for its clarity and simplicity\footnote{Another way would be to make the learning rate $\alpha$ stochastic, instead of the cost. \citet{mazzola1996bayesian} considered how a Bayesian learner benefits from more production not only by decreasing costs, but also by improved estimates of the learning parameter.}. The cost of production at $t=1$, interpreted as the average (or constant) within-period cost, is then given by
\begin{eqnarray}
c_1 & = & c_0 \left( \frac{z_0}{z_1} \right)^{\alpha} e^{\eta} =  c_0 \left( \frac{z_0}{z_0 + q} \right)^{\alpha} e^{\eta} \label{cost}.
\end{eqnarray}
So there is a distribution of possible future costs $c_1$, and Eq.\ (\ref{cost}) shows clearly how it depends on: \emph{i}) the current state, $c_0$, $z_0$, of the technology\footnote{Note that it is the presence of $z_0$ here that distinguishes between learning effects and increasing returns to scale.}, \emph{ii}) the technology's experience exponent $\alpha$, \emph{iii}) the choice of production $q$ over the period, and \emph{iv}) the noise distribution $\eta$.

Next, we suppose that the shock is normally distributed with mean zero\footnote{Nonzero mean noise is discussed in Section \ref{section:two_period_model}.} and variance $\sigma^2$, $\eta  \sim \mathcal{N}(0,\sigma^2)$.
This noise model is known to be a reasonable assumption for financial assets with lognormal returns, but some justification is required when considering technologies.
\citet{lafond2018} found that this model gave a reasonably good fit to data on 51 technology time series, in the sense of predicting theoretical forecast errors in line with realised forecast errors, although their preferred model allows for autocorrelation.

Thus cost is log-normally distributed, and by standard log-normal properties its expectation and variance are given by
\begin{equation}
\label{cost_expectation} 
\mathbb{E} \left[ c_1 \right] = c_0 \left( \frac{z_0}{z_0 + q} \right)^{\alpha} e^{ \sigma^2 / 2},
\end{equation}
\begin{equation}
\label{cost_variance}
\text{Var} \left( c_1 \right) = c_0^2 \left( \frac{z_0}{z_0 + q} \right)^{2 \alpha} e^{ \sigma^2} \left( e^{ \sigma^2} -1 \right).
\end{equation}
These two properties of the stochastic experience curve, specified uniquely by the four parameters $c_0, z_0, \alpha, \sigma$, will now be used to construct the portfolio model.

\subsection{The optimization}

Consider two independent technologies, $A$ and $B$, each evolving according to the form of Wright's law proposed above, with their own technology-specific parameters. We label variables and parameters with superscripts (e.g.\ $q^A, c_0^A, z_0^A, \alpha^A, \sigma^A$). Suppose the technologies are perfect substitutes\footnote{While the perfect substitutability assumption is essential in this model, in reality technologies are often not continuously varying substitutes, and it may not be possible to adopt just a bit of several different technologies. Indeed, many technology adoption decisions are entirely binary, such as the choice of firm-wide software systems. This is a limitation of the model, and the domain of application should therefore be chosen carefully.} and that there is a fixed, exogenous demand $K$, which must be satisfied exactly by some combination of production of the two technologies\footnote{Since demand and total production are assumed equal throughout we use the terms interchangeably. It is assumed that the demand is inelastic and prices are determined competitively (as is typical in energy markets). Under these conditions cost minimization is equivalent to profit maximization.}, i.e.\ there is a production constraint $K = q^A + q^B$.
Production is non-negative, so $q^A, q^B \in [0,K]$, and choosing $q^A$ also determines $q^B = K - q^A$. We use $q^A$ as the control variable in the following optimization and present results in terms of the share of total production in technology $A$, $q^A/K$. Let the total cost of production during the period be $V(q^A)$. This is just the sum of unit costs times units produced
\begin{equation}
V(q^A) = \sum_{i=A,B} c_1^i q^i,
\label{total_cost}
\end{equation}
where stochastic costs $c_1^i$ depend nonlinearly on productions $q^i$, as in Eq.\ (\ref{cost}). Thus for a fixed, known set of technology parameters $\{c_0^i, z_0^i, \alpha^i, \sigma^i\}_{i=A,B}$ and total demand $K$, each choice of production $q^A$ maps to a distribution of total costs $V$. The tools for addressing this type of problem are well developed, see for example \citet{Krey2013}. The goal here is to understand how the parameters and the choice of production together generate the total system cost distribution, from which an optimal production portfolio may be identified. We perform a mean-variance analysis on $V$ because it is simple, intuitive and illustrates clearly the key features of the system\footnote{Since empirical technology cost noise shocks are found to fit a lognormal distribution fairly well, as discussed previously, standard results from the finance literature apply here. In particular, use of the mean-variance decision framework in the one-period setting is justified as it provides a good approximation to all commonly used utility functions (\citet{pulley1981_meanvariance}, \citet{kroll1984_meanvariance}).
However, the choice of utility function in a multi-period setting (as we consider in Section \ref{section:two_period_model}) is much more subtle, and a different objective function may be preferable.}. Let $\lambda \geq 0$ be a risk aversion parameter and $f$ be the mean-variance objective function. The optimization problem is then
\begin{eqnarray}
\label{obj_fn_simplest_form}
&\underset{q^A}{\text{minimize:}} & f(q^A) = \mathbb{E} \left[ V(q^A) \right] + \lambda \text{Var} \left( V(q^A) \right) \\[0.3cm]
&\text{subject to:}& q^A \in [0, K]. \nonumber
\end{eqnarray}
The aim therefore is to find the production mix which, while meeting the production constraint, minimizes the expected total cost of production, plus an additional term characterizing the spread of the distribution of possible outcomes. The risk aversion parameter $\lambda$ scales the contribution of the variance term in $f$, reflecting the extent to which the decision maker prefers to minimize exposure to cost uncertainty. In the risk-neutral case ($\lambda=0$) the variance term has zero weight so the optimization just discovers the production mix with lowest expected total cost (in this case just a single technology). Conversely, in the high risk aversion case ($\lambda \gg 1$) the second term in $f$ dominates the first and so the optimization discovers the production mix with lowest total cost uncertainty, regardless of its expectation. In the intermediate regime both terms play a significant role in determining the outcome of the optimization.
Using Eqs.\ (\ref{cost_expectation}), (\ref{cost_variance}) and (\ref{total_cost}) the objective function in problem (\ref{obj_fn_simplest_form}) may be written explicitly as
\begin{eqnarray}
f(q^A) & = & \sum_{i = A, B} c_0^i \left( \frac{z_0^i}{z_0^i + q^i} \right)^{\alpha^i} e^{ (\sigma^i)^2 / 2} q^i \nonumber \\
& & + \lambda \left( c_0^i \left( \frac{z_0^i}{z_0^i + q^i} \right)^{\alpha^i} q^i \right)^2 e^{ (\sigma^i)^2} \left( e^{ (\sigma^i)^2} -1 \right).
\label{obj_fn_explicit}
\end{eqnarray}
Thus $f$ is just the sum of one cost-expectation-based component and one cost-variance-based component for each technology; covariance terms are zero due to the technology independence assumption (i.e.\ $\eta^A$ and $\eta^B$ are uncorrelated). (The case of correlated noise is considered in Section \ref{subsection:correlations}.)

This is a non-convex optimization problem so it may have more than one local minimum. Since there is only one free variable though, $q^A$, it is relatively quick to solve by brute force optimization. Denote the optimum by $q_*^A$

Despite the simplicity of the model, the scope for understanding its behaviour via standard analytical techniques is rather limited.
This is because the product terms $(z_0^i + q^i)^{-\alpha^i} q^i$ in the objective function mean that differentiation of $f$ just generates more and more similar product terms, which makes closed-form expressions for optima or other system properties only possible in a few restricted cases.
Most of our results and analysis are therefore based on numerical optimization (and so were checked extensively to ensure they are representative of the whole parameter space).

\subsection{Technological maturity and the no-learning limit}
\label{section:maturity}

\subsubsection{Markowitz portfolios}
\label{markowitz}
Consider briefly the topic of Markowitz portfolio analysis for standard financial assets \citep{markowitz1952portfolio}. Let $\mathbf{r} = (r_1, \ldots, r_n)^T$ be a vector of stochastic returns (possibly correlated) and $\mathbf{w} = (w_1, \ldots, w_n)^T$ be a vector of portfolio weights. The portfolio return distribution is $V(\mathbf{w}) = \mathbf{w}^T \mathbf{r}$, on which a mean-variance optimization is carried out, with $\mathbf{w}$ as control variable. The classic form of the problem is
\begin{eqnarray}
\label{markowitz_classic}
&\underset{\mathbf{w}}{\text{maximize:}} & f(\mathbf{w}) = \mathbb{E} \left[ V(\mathbf{w}) \right] - \lambda \text{Var} \left( V(\mathbf{w}) \right) \\[0.3cm]
&\text{subject to:}& \sum_{j=1, \ldots, n} w_j = 1. \nonumber
\end{eqnarray}
Since this is a mean-variance optimization it looks very similar to our technology portfolio problem (\ref{obj_fn_simplest_form}). There are several differences though; three are superficial but one is fundamental.

First, in the Markowitz case the decision maker seeks high expected portfolio return and low variance, while in the technology case they seek low expected portfolio cost \textit{and} low variance\footnote{This is also the case in the optimal liquidation problem, see e.g.\ \citet{almgren2001optimal}.} --- hence the sign difference of the variance terms in (\ref{obj_fn_simplest_form}) and (\ref{markowitz_classic}).
Second, short-selling is in general allowed, so portfolio weights $w_j$ are not restricted to being non-negative. 
Third, returns are generally assumed to be correlated, and a lot of attention is paid to understanding these correlations.
Finally though, the fundamental difference between the two problems is that in the Markowitz case asset returns are purely stochastic, so portfolio weights do not affect asset performance, while in the experience curve model the stochastic costs depend explicitly on production, so portfolio weights \textit{do} affect technology performances. The more one invests in a given technology the better it gets, on average; there is nonlinear feedback in the technology portfolio model but not in the Markowitz model.

\subsubsection{Comparing financial and technology portfolios}
\label{markowitz_comparison}

To better understand the differences between the two portfolio types, we make a more accurate comparison by using a restricted version of the Markowitz model: the no short-selling, enforced budget, uncorrelated, two-asset model. This is a direct equivalent of our technology portfolio problem in a standard financial setting. It eliminates the second and third superficial differences listed above, making it easier to observe feedback effects.

Suppose there are two assets, $A$ and $B$, with uncorrelated normal returns $r^A \sim \mathcal{N}(\mu^A, (s^A)^2)$ and $r^B \sim \mathcal{N}(\mu^B, (s^B)^2)$. Then let $q^A$ and $q^B = (1 - q^A)$ be the proportion of wealth invested in $A$ and $B$ respectively, with $q^A, q^B \in [0,1]$ (the no short-selling condition). The portfolio return distribution is then 
$V(q^A) = \sum_{i=A,B} r^i q^i$, and the objective function to be maximized is
\begin{eqnarray}
f(q^A) &=& \mathbb{E} \left[ V(q^A) \right] - \lambda \text{Var} \left( V(q^A) \right) \label{markowitz_obj_fn_eq} \\
&=&  \sum_{i=A,B} \mu^i q^i  - \lambda \left( s^i q^i \right)^2. \label{markowitz_obj_fn_eq2}
\end{eqnarray}
Note that this is quadratic in portfolio weights $q^i$.
Then returning to the technology portfolio problem and considering the role of demand $K$ and initial cumulative productions $z_0^A$ and $z_0^B$ in the objective function, a simple calculation reveals the connection between the financial and technology models. 
Observe that the technologies objective function, Eq.\ (\ref{obj_fn_explicit}), may be written
\begin{equation}
f(q^A) =  \sum_{i=A,B} \frac{c_0^i q^i}{(1 + \frac{q^i}{z_0^i} )^{\alpha^i}} e^{ (\sigma^i)^2 / 2}  + \lambda \left( \frac{c_0^i q^i}{(1 + \frac{q^i}{z_0^i} )^{\alpha^i}} \right)^2 e^{ (\sigma^i)^2} \left( e^{(\sigma^i)^2} -1 \right).
\label{expandable_f}
\end{equation}
When $q^i/z_0^i$ is small we can approximate this in a simpler form.
If the maximum future production of technology $i$ is much less than its current cumulative production ($K \ll z_0^i$), then $q^i / z_0^i \ll 1$, and the binomial series representation
\begin{equation}
(1 + \frac{q^i}{z_0^i} )^{-\alpha^i} = 1 - \alpha^i \frac{q^i}{z_0^i} + \alpha^i (\alpha^i+1) \left( \frac{q^i}{z_0^i} \right)^2 + \ldots
\label{param_expansion}
\end{equation}
may be used. Thus if $K \ll z_0^i$ for both technologies then to zeroth order the objective function may be approximated as
\begin{equation}
f(q^A) \approx \sum_{i=A,B} c_0^i e^{ (\sigma^i)^2 / 2} q^i + \lambda \left( c_0^i \right)^2 e^{ (\sigma^i)^2} \left( e^{(\sigma^i)^2} -1 \right) (q^i)^2,
\label{markowitz_approx}
\end{equation}
which no longer includes the experience exponents $\alpha^i$. Appendix \ref{appendix:expansion} shows details of the expansion, plus higher order terms.
Apart from the sign difference of the variance component, this has the same form as the Markowitz model (Eq.\ (\ref{markowitz_obj_fn_eq2})), i.e.\ it is quadratic in production\footnote{Note that the $\sigma^i$ terms remain in the expectation component here due to the particular noise model used (Eq.\ (\ref{cost_expectation})). They are fixed and independent of $q^i$ (and indeed could be avoided with a different choice of noise), so do not affect the argument.}.
In this limit learning plays no part, and there is no feedback process by which production affects future costs (since this is represented by the higher order terms). Hence a Markowitz-like portfolio problem is the limiting case of the Wright's law portfolio problem as learning effects tend to zero.

Eq.\ (\ref{param_expansion}) shows that a low-learning regime can exist in two ways for a given technology: first, if its learning rate is intrinsically small, and second, if its initial cumulative production is very large compared to the total demand. The latter condition is problem-specific, since it depends on $K$, not just on the technology itself. All else being equal, as $K \rightarrow 0$ technologies behave increasingly like standard financial assets, as noise increasingly dominates learning effects. Furthermore, very mature technologies automatically behave like standard financial assets in the model (since the incremental gains due to learning decrease with maturity by definition in Wright's law). Note that this analysis relies on the assumption that model parameters are static, e.g.\ experience exponents are constant and do not depend on the size of $K$. This assumption would require justification in any practical application, and indeed is closely related to the question of whether a single- or multi-period framework is more appropriate (the latter could allow for a more fine-grained approach to modelling technological maturity, for example).
Nevertheless the simple analytical connection between the two portfolio systems shown here is of note, as it reveals an interesting perspective on technological maturity in a portfolio setting.

Within any given problem then, each technology lies somewhere on a spectrum between more \textit{technology-like} and more \textit{asset-like}, depending on the entire set of parameters. We use ``asset-like'' simply to mean that learning effects are negligible relative to noise, as with standard financial assets.

Finally, consider how the learning and non-learning portfolio problems differ analytically at lowest orders. As shown in Appendix \ref{appendix:expansion}, the approximation to $f$ including the lowest order ``learning'' terms (i.e.\ terms in $- \alpha^i \frac{q^i}{z_0^i}$) is
\begin{equation}
f(q^A) \approx \sum_{i=A,B} c_0^i e^{ (\sigma^i)^2 / 2} \left( 1 - \alpha^i \frac{q^i}{z_0^i} \right) q^i + \lambda \left( c_0^i \right)^2 e^{ (\sigma^i)^2} \left( e^{(\sigma^i)^2} -1 \right) \left( 1 - 2 \alpha^i \frac{q^i}{z_0^i} \right) (q^i)^2.
\label{markowitz_approx_first_order}
\end{equation}
Thus the most straightforward effect of learning is to reduce both the expectation and variance components linearly in $\alpha^i$, so that the technology with higher $\alpha^i$ will perform relatively better in the optimization. In addition though, observe that while the zeroth-order approximation to $f$ (Eq.\ (\ref{markowitz_approx})) is quadratic in $q^A$, and hence always has just one single minimum, the first-order approximation to $f$ is cubic in $q^A$, and may therefore have two local minima inside the optimization range (depending on parameters). The introduction of learning therefore corresponds to the introduction of multiple local optima of the objective function.

\section{Optimization results}
\label{section:t1}
The goal here is to understand how the optimal allocation of production between the two competing technologies depends on the technology-specific learning parameters (experience exponent $\alpha$ and volatility $\sigma$) and the initial conditions (cost competitiveness $c_0$ and cumulative production $z_0$) under varying levels of risk aversion $\lambda$, for fixed demand $K$.
To do this we first hold all model parameters constant, then vary technology $B$ experience exponent $\alpha^B$ and risk aversion $\lambda$. This generates a grid of tuples $(\alpha^B, \lambda)$. At each point of this grid the optimization (\ref{obj_fn_simplest_form}) is performed, and the resulting collection of optima is plotted, giving the surface of optimal production of technology $A$ as a share of total production, ${q_*^A}/K$. The whole process may then be repeated for each of the other technology parameters $\sigma^B$, $c_0^B$ and $z_0^B$.

\subsection{Effects of experience exponents $\alpha$}
\label{subsection:effects_of_exp_exp}

We set the initial conditions and parameter values to those shown in Table \ref{alpha2-lambda-surface-param-values}. Almost identical technologies are used here as this allows us to understand the effects of varying different parameters most effectively. (Asymmetrical technologies are considered in Section \ref{section:lockin}.)
Note that the total demand is twice the initial cumulative production of each technology. Hence the technologies are relatively immature, in the sense that there is plenty of potential left for learning to take place relative to how much has occurred in the past.
As shown above, this is necessary since if both technologies are sufficiently mature a nearly-Markowitz scenario emerges.

\begin{figure}[H]
  \centering
  \includegraphics[width=1.0\textwidth]{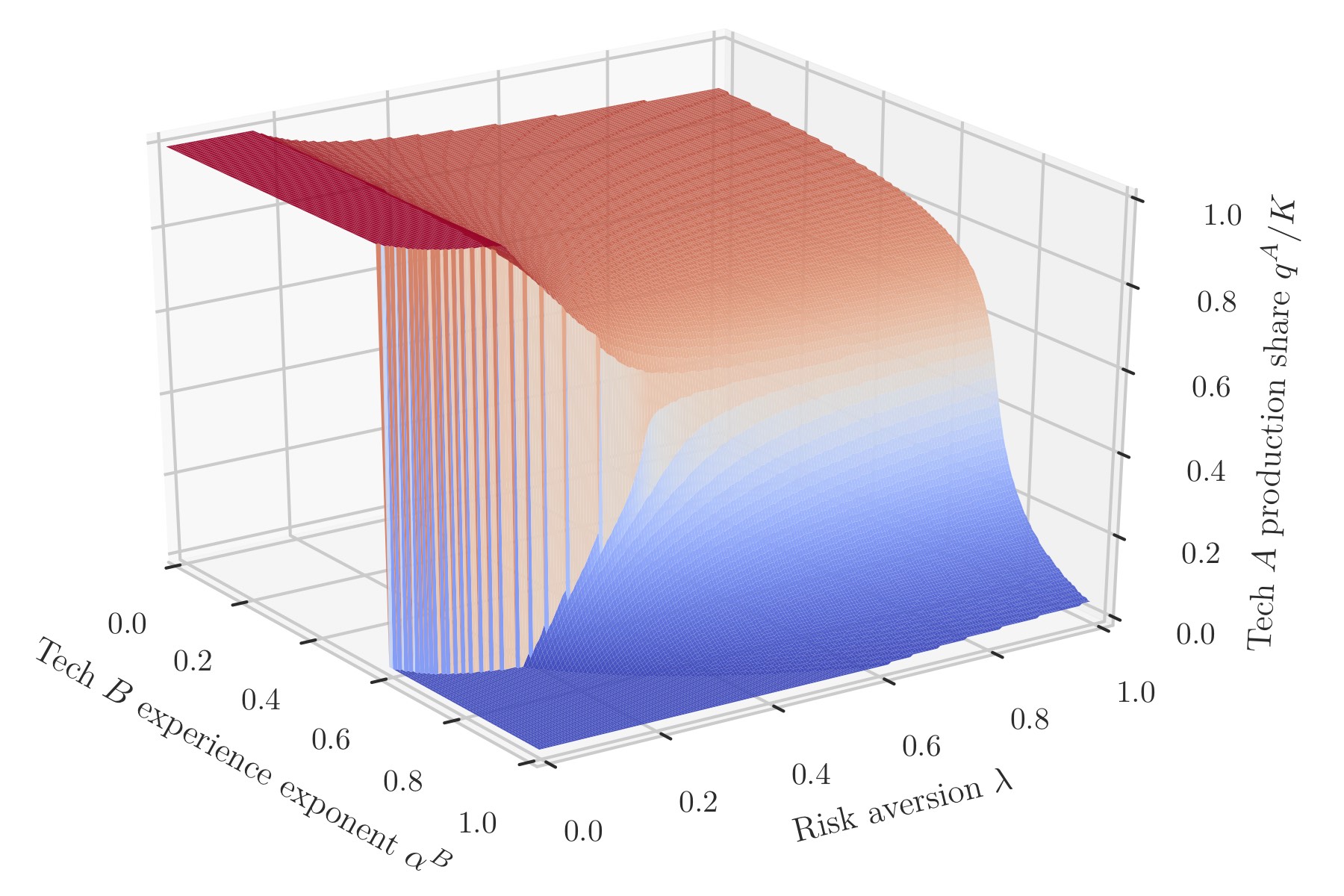}
  \caption{Surface of the optimal production in technology $A$ as a share of total production. This shows how the optimal portfolio varies with risk aversion $\lambda$ and technology $B$ experience exponent $\alpha^B$ (with $\alpha^A$ fixed at 0.5). Red areas correspond to higher production of technology $A$ being optimal, and blue areas to higher production of technology $B$ being optimal.
Low risk aversion leads to more specialized portfolios and greater parameter sensitivity (represented by the surface discontinuity), while high risk aversion leads to greater diversification and lower parameter sensitivity.
Parameter values are shown in Table \ref{alpha2-lambda-surface-param-values}.}
  \label{alpha2-lambda-surface}
\end{figure}

Fig.\ \ref{alpha2-lambda-surface} shows the surface of optimal technology $A$ production share, ${q_*^A}/K$, over a grid of $\alpha^B$ and $\lambda$ values. This shows how risk aversion and relative experience exponents affect the composition of the optimal portfolio.
\begin{table}[H]
\centering
\begin{tabular}{|clcc|}
\hline
Symbol & Description & Tech $A$ & Tech $B$\\
\hline
$z_0$ & Technology maturity & 1 & 1\\
$c_0$ & Initial cost & 2 & 2\\
$\alpha$ & Experience exponent & 0.5 & [0-1] \\
$\sigma$ & Technology volatility & 1.0 & 1.1\\
\hline
$K$ & Demand &\multicolumn{2}{c|}{2} \\
$\lambda$ &Risk aversion&\multicolumn{2}{c|}{[0-1]} \\
\hline
\end{tabular}
\caption{Parameter values for the case of two almost identical technologies, as used in Fig.\ \ref{alpha2-lambda-surface}}
\label{alpha2-lambda-surface-param-values}
\end{table}
When risk aversion is low the optimal strategy is to concentrate production entirely in either $A$ or $B$ (the dark red and blue plateau regions), depending on relative experience exponents. When risk aversion is high portfolios are diversified over both technologies. This is consistent with a general understanding of both deterministic experience curves (in which specialization is always optimal) and standard portfolio theory (in which diversification reduces portfolio risk). However, the nature of the transitions between these regimes depends on model parameters and is of great interest. For low to moderate risk aversion there is a discontinuity in the surface, indicating a region of extreme sensitivity to model parameters. In this region an incremental change in either experience exponent or risk aversion can lead to a large change in the optimal portfolio. In contrast, for high risk aversion the surface is smooth, so the optimal portfolio is robust to small changes in parameters. As we shall see (in Section \ref{local_mins_section}), this is caused by the existence of multiple local minima of the objective function in the low risk aversion regime, and a single global minimum in the high risk aversion regime.

On the $\lambda=0$ boundary, variance terms do not feature in the optimization so production is concentrated in the technology with the best expected outcome. As risk aversion increases, up to around 0.2, the asymmetry in noise variance becomes apparent and the threshold for switching from 100\% $A$ to 100\% $B$ gradually shifts to larger $\alpha^B$ values. The preference for the higher experience exponent technology ($B$ in this region) is traded off against a preference for the less noisy technology ($A$), since the optimization penalizes higher noise variance. As risk aversion increases further portfolios become increasingly balanced. The surface discontinuity becomes less pronounced as the two local optima on either side of it approach a common value. Eventually the discontinuity disappears, and a single stable global optimum exists thereafter.
(Only $\lambda=0$ is a strict boundary in the model, and the surface extends in the other directions beyond the bounds shown.)

Therefore some combinations of technologies and risk preferences are more robust than others: in some regions the solution is not particularly sensitive to changes in the underlying parameters, while for others it is extremely sensitive. In the unstable regions, a parameter estimation error could lead to a mix of technologies being chosen that is very far from the true optimal mix.

\subsection{Comparison with Markowitz portfolios}
\label{subsection:markowitz}

\begin{figure}
  \centering
  \includegraphics[width=1.0\textwidth]{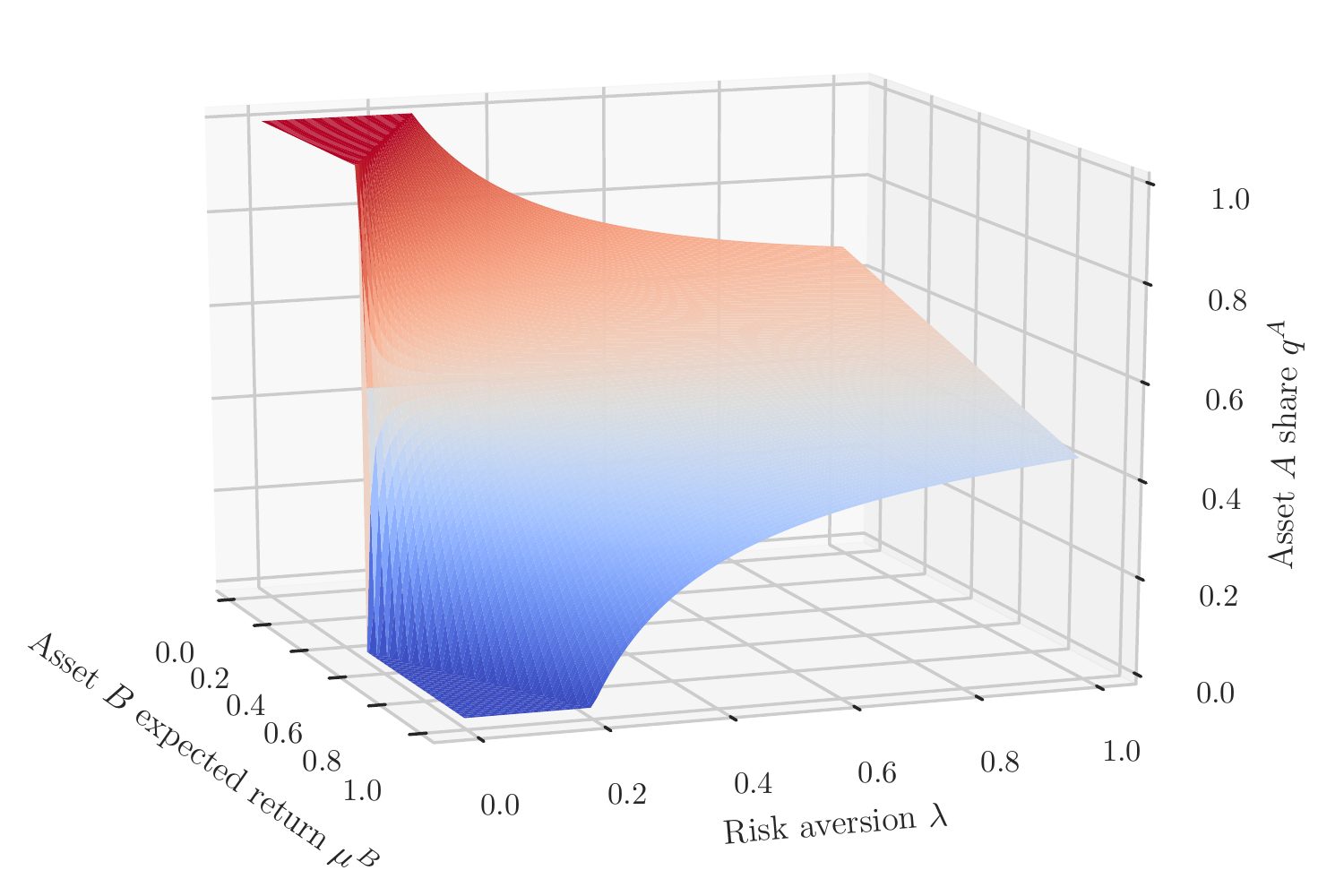}
  \caption{The Markowitz portfolio analogue of the technology portfolio surface shown in Fig.\ \ref{alpha2-lambda-surface}. This is the surface of optimal investment share in asset $A$ for varying values of risk aversion and asset $B$ expected return. Portfolios are more diversified for high risk aversion and more specialized for low risk aversion as before, and there still exist regions of full specialization, in which one technology sufficiently outperforms the other. However, in contrast to the case of technologies the surface is continuous $\forall \lambda>0$, due to the convexity of the objective function.  }
  \label{markowitz_plot}
\end{figure}

To illustrate how nonlinearities in the technology portfolio affect the optimization results relative to the financial assets case, we plot the corresponding surface of optimal portfolio weights for the equivalent Markowitz system, Eq.\ (\ref{markowitz_obj_fn_eq}).
With model parameters in Eq.\ (\ref{markowitz_obj_fn_eq2}) set to $\mu^A=0.5$, $s^A=1.0$, $s^B=1.1$, Fig.\ \ref{markowitz_plot} shows the surface of optima over a grid of varying asset $B$ expected return $\mu^B$, and risk aversion $\lambda$.
The usual patterns are present: portfolios are more diversified for higher risk aversion and more specialized for lower risk aversion. Full specialization occurs when one asset sufficiently outperforms the other, again giving the dark red and blue plateau regions. However the crucial difference is that now the surface is continuous everywhere except at the single point on the $\lambda=0$ boundary where $\mu^A = \mu^B$. There are no positive values of risk aversion at which portfolios transition instantaneously from one state to another; portfolios vary continuously with both risk aversion and model parameters. This is because the Markowitz problem is convex. Without the Wright's law nonlinearity in $f$ there do not exist multiple local minima for portfolios to instantaneously switch between as parameters vary, and hence no unstable regions of parameter space.

\subsection{Analysis}
\label{section:analytics}
Next we present some analytical observations which help in understanding the character of the problem and the shape of the surface in Fig.\ \ref{alpha2-lambda-surface}.

\subsubsection{Corner and interior solutions}
\label{boundarysolutions}
Since the optimization domain is bounded ($q^A \in [0,K]$), solutions are either corner solutions or interior solutions. Corner solutions ($q_*^A=0 \text{ or } K$) satisfy $f'(q_*^A) \neq 0$ almost everywhere in parameter space, while interior solutions ($q_*^A \in (0,K)$) always satisfy $f'(q_*^A)=0$. Corner solutions form both the dark red horizontal plateau with $q_*^A = K$ on the left of Fig.\ \ref{alpha2-lambda-surface} and the dark blue horizontal floor section with $q_*^A  =  0$ at the front of the plot (plus the equivalent areas on Fig.\ \ref{markowitz_plot}). All other points of the surface are interior solutions, at which optimal portfolios are diversified.

\subsubsection{Local and global minima of the objective function}
\label{local_mins_section}
\begin{figure}
  \centering
  \includegraphics[width=1.0\textwidth]{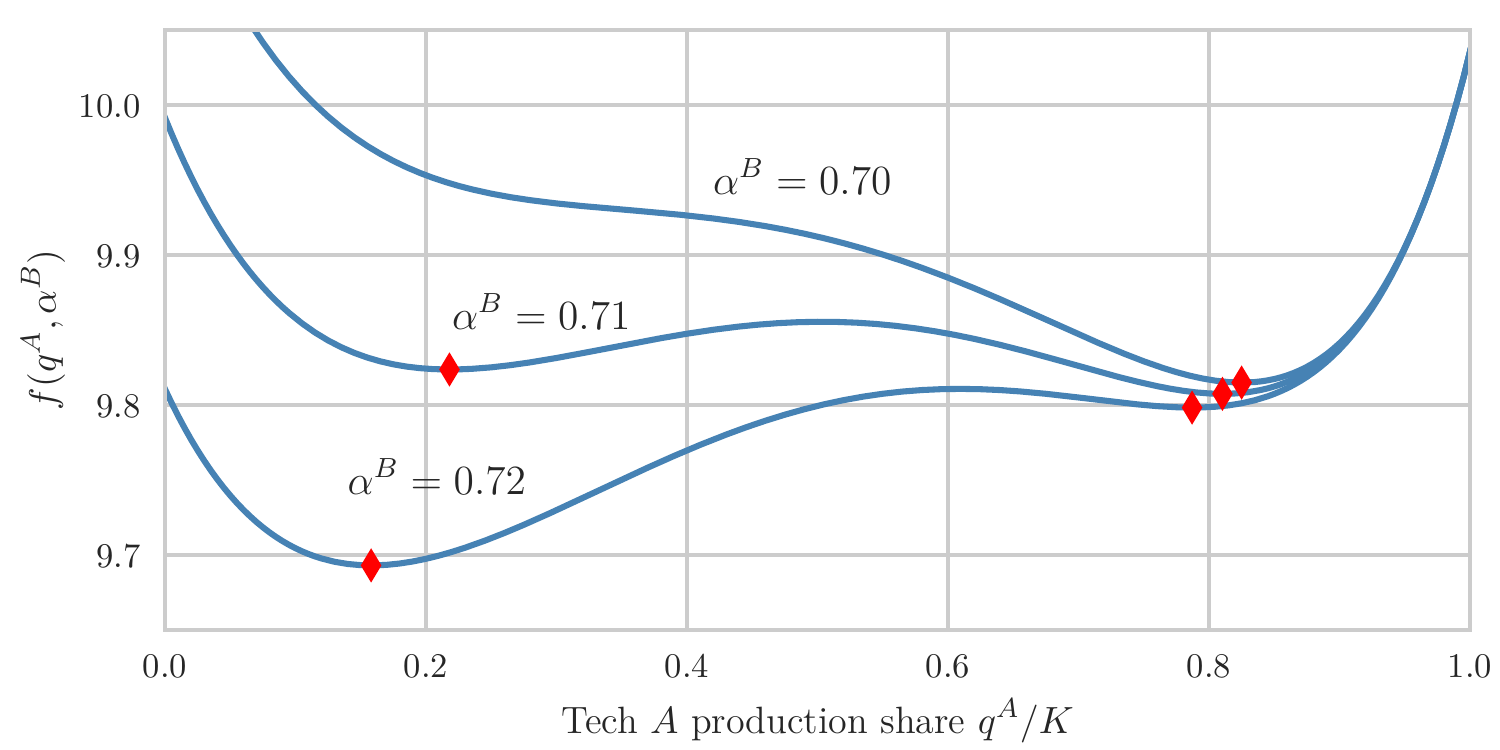}
  \caption{The objective function for three different technology $B$ experience exponents (emphasized by writing $\alpha^B$ as an argument of $f$ here). Minima are shown in red, risk aversion is fixed at $\lambda = 0.25$ and all other parameters are as before. For smaller $\alpha^B$ there is a single interior local minimum with production concentrated mainly in $A$. As $\alpha^B$ increases a second local minimum appears, which then becomes the global minimum, and production switches to being mainly concentrated in $B$. This is what happens as the surface discontinuity in Fig.\ \ref{alpha2-lambda-surface} is crossed --- highly differentiated portfolios of approximately equal objective value exist simultaneously.}
  \label{obj-fn-plot1}
  \centering
  \includegraphics[width=1.0\textwidth]{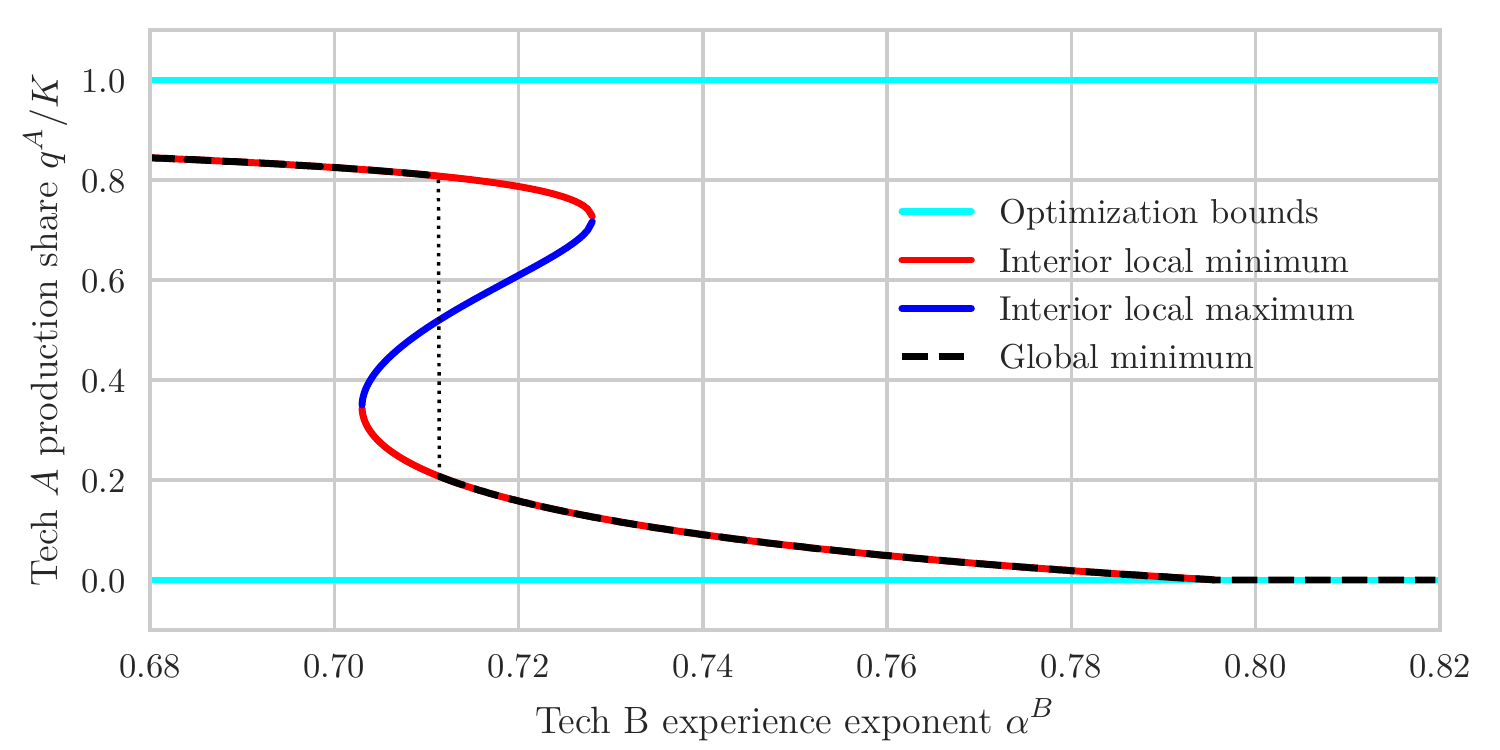}
  \caption{Locations of the optima of the objective function for varying $\alpha^B$, corresponding to Fig.\ \ref{obj-fn-plot1}. This is the $\lambda = 0.25$ section through Fig.\ \ref{alpha2-lambda-surface}. Distinct local minima emerge and disappear as $\alpha^B$ varies. At the critical value $\alpha^B_{\textit{switch}} \approx 0.71$ the global optimum switches instantaneously between the two minima.}
  \label{cusp_slice_alphaB}
\end{figure}

The nonlinearity in the model generates interesting behaviour because in some regions of parameter space the objective function has multiple local optima. Fig.\ \ref{obj-fn-plot1} shows how the objective function varies along one particular line in parameter space: risk aversion is fixed at $\lambda=0.25$ and technology $B$ experience exponent is varied (so this corresponds to a section through Fig.\ \ref{alpha2-lambda-surface}).
The objective function is plotted for three different values of $\alpha^B$, showing how distinct local minima emerge and disappear. As $\alpha^B$ varies the global minimum switches from one local minimum to another, and very different portfolios of approximately equal objective value exist simultaneously. When the surface discontinuity in Fig.\ \ref{alpha2-lambda-surface} is crossed the global minimum switches from one local minimum to the other. This means that a parameter estimation error could lead to a portfolio significantly different to the correct optimal portfolio being chosen. Fig.\ \ref{cusp_slice_alphaB} plots the locations of the different optima against $\alpha^B$. This shows how, if the measured value of $\alpha^B$ is, for example, $0.7 \pm 0.02$, then the optimal production share is roughly a 20:80 split, but either technology could be the dominant one, depending on what the true value really is.

Finally, since Fig.\ \ref{cusp_slice_alphaB} is just the $\lambda=0.25$ section through Fig.\ \ref{alpha2-lambda-surface}, it is apparent that if all optima were plotted on Fig.\ \ref{alpha2-lambda-surface}, not just the global minima, the surface would double back under itself in a fold, smoothly connecting the upper and lower edges of the discontinuity. This type of geometry is well-known from the cusp catastrophe bifurcation (see e.g.\ \citet{zeeman1976catastrophe}, \citet{poston2014catastrophe}). Although our setting is different, since parameters here are not dynamic, the similarity is worth noting; both involve plotting the zeros of an underlying nonlinear system, resulting in a multivalued surface representing alternative stable states. Appendix \ref{appendix:technical_points} describes some basic properties of the system that can be derived analytically.

\subsubsection{Effects of other technology parameters $c_0$, $z_0$ and $\sigma$}

\begin{figure}
  \centering
  \includegraphics[width=1.0\textwidth]{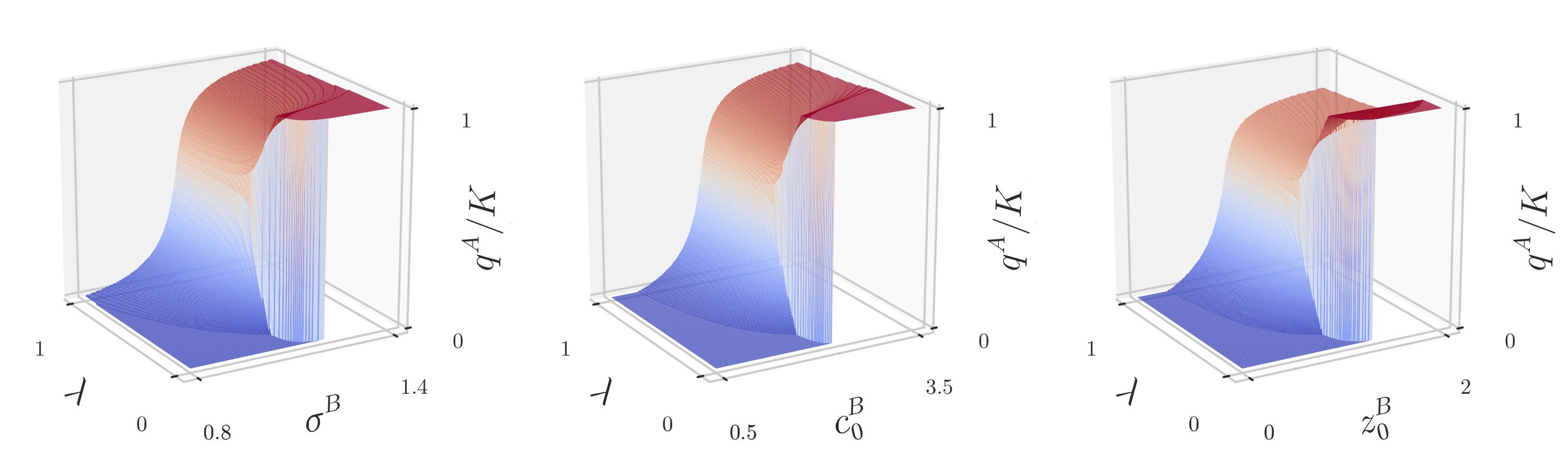}
  \caption{Surfaces of optimal technology $A$ production share, analogous to Fig.\ \ref{alpha2-lambda-surface}, but for varying risk aversion $\lambda$ and technology $B$ parameters $\sigma^B, c_0^B, z_0^B$.}
  \label{three_risk_surfaces}
\end{figure}

Of the four technology parameter pairs $\{c_0^i, z_0^i, \alpha^i, \sigma^i\}_{i=A,B}$, so far we have only studied one: the experience exponents $\alpha^i$. To do this we held all parameters constant, including $\alpha^A$, then solved the optimization over a grid of varying $\alpha^B$ and $\lambda$ values. To study the other three parameter pairs the same procedure is repeated for each of them in turn.
Fig.\ \ref{three_risk_surfaces} shows the results, with $\alpha^B=0.65$ and all other fixed parameters set to the values in Table \ref{alpha2-lambda-surface-param-values}.
The plots appear reversed relative to Fig.\ \ref{alpha2-lambda-surface} because while higher values of $\alpha^i$ correspond to better performance in the model, the opposite is true for the other parameters (e.g.\ higher values of $\sigma^i$ are penalised more).
Thus we see that all technology parameters produce the same effect as the experience exponents, each pair having its own distinct regions of stability and instability in parameter space.
This makes sense intuitively by considering Fig.\ \ref{obj-fn-plot1} --- at any point in parameter space the objective function is a curve similar to these, and perturbing \emph{any one} of the underlying parameters will cause a similar smooth change in the curve, and its optima, just as it does for the $\alpha^i$.
The only remaining parameter in the model is demand $K$, which we examine next.

\subsection{Effect of total demand; demand-driven lock-in}
\label{demand_section}
In the example system used so far, total demand $K$ is twice the initial cumulative production of both technologies ($K = 2 z_0^A = 2 z_0^B$). The potential for learning is therefore high and the model behaves very differently than the Markowitz case.
We now consider how this behaviour changes as demand is varied.
Intuitively we would expect that, all else being equal, the larger $K$ is, the more potential there is for experience to accrue, so the more concentrated the portfolio will be in one technology.
We demonstrate that our model produces this behaviour, and show in detail the transition from the small-$K$, low-learning regime to the large-$K$, high-learning regime.

\begin{figure}
  \centering
  \includegraphics[width=1.0\textwidth]{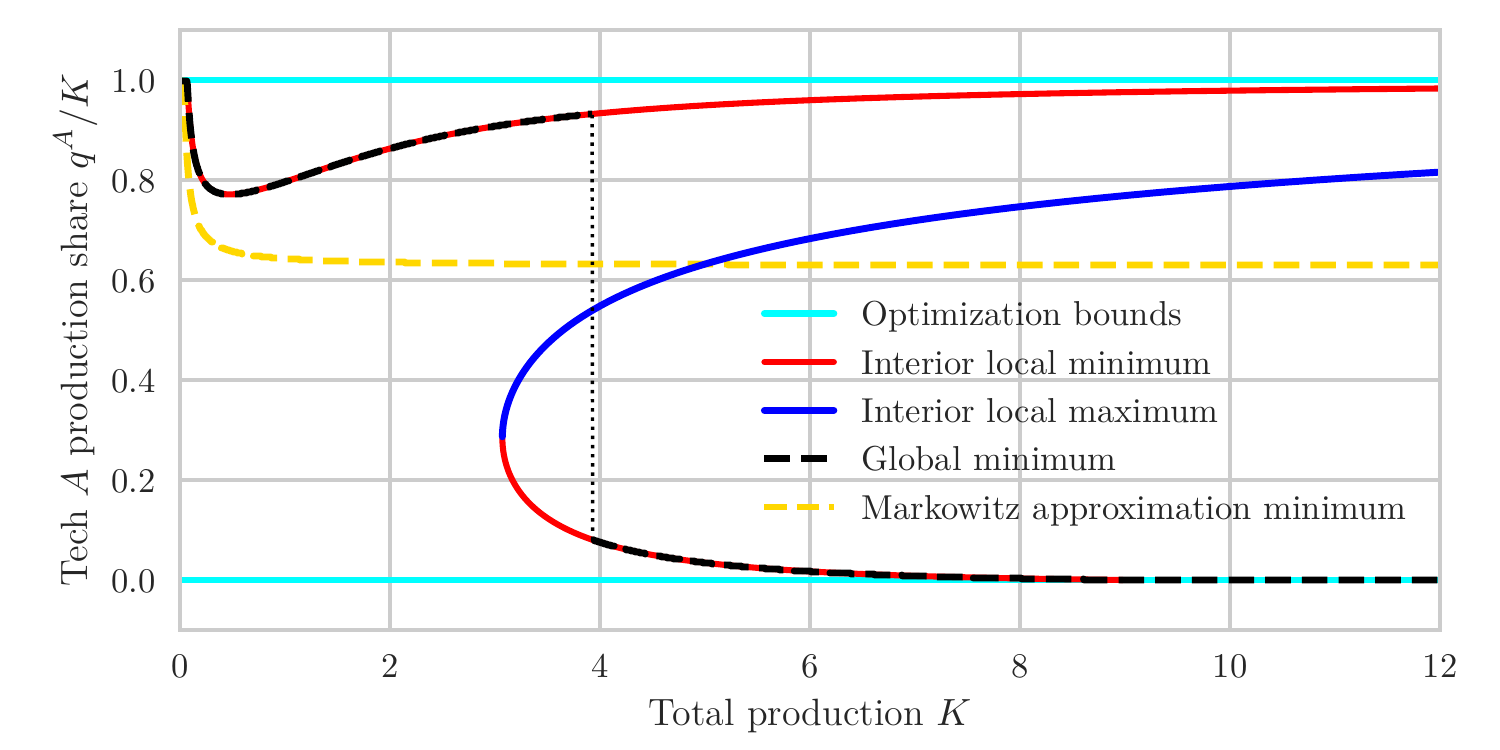}
  \caption{Plot showing how optima of the objective function vary with total production $K$. The minimum of the Markowitz approximation to   $f$ (Eq.\ \ref{markowitz_approx}) is also shown for comparison. As $K$ goes from approximately 0 to 1 the system transitions from a Markowitz-like, low-learning regime to a technology-like, learning regime. When demand is small (approximately $K<4$) the more certain, slower progressing technology $A$ ($\alpha^A=0.5$) dominates the portfolio, but when demand is high there is enough scope for progress to occur that the noisier, faster progressing technology $B$ ($\alpha^B=0.65$) becomes optimal. The transition between these states is instantaneous as the global optimum switches between different local minima of equal objective value. Risk aversion is fixed at $\lambda=0.25$ and other parameter values are those shown in Table \ref{alpha2-lambda-surface-param-values} as before.}
  \label{cusp_slice_K}
\end{figure}

In Fig.\ \ref{cusp_slice_K}, technology $A$ experience exponent is still fixed at $\alpha^A = 0.5$, and instead of varying $\alpha^B$ as before we also fix it, at $\alpha^B = 0.65$, and vary $K$.
Risk aversion is fixed at $\lambda = 0.25$ and other parameters are those shown in Table \ref{alpha2-lambda-surface-param-values} as before (so technology $B$ progresses faster than $A$ but is still slightly noisier).
The plot shows how the local and global optima of the objective function vary with $K$ while all other parameters are held constant --- it is the analogue of Fig.\ \ref{cusp_slice_alphaB}, but with $K$ as the independent variable.
As well as the maximum and minima of $f$, the plot also shows the minimum of the Markowitz approximation to $f$ for comparison.

Again we observe instantaneous switching between portfolio states as demand varies.
As $K$ goes from approximately 0 to 1 the share of technology $A$ in the optimal portfolio (dashed black line) initially decreases sharply then reverses direction and increases again, due to the increased potential for learning.
Here the technologies transition from behaving in a more ``asset-like'' way to a more ``technology-like'' way, as more higher order terms in the series expansion of $f$ start to have an impact (see Eq.\ (\ref{markowitz_approx_first_order})).
Indeed, for very small $K$, the minimum of $f$ and the minimum of the Markowitz approximation to $f$ roughly coincide (i.e.\ the black and yellow dashed lines), and the solution for technologies is almost the same as for financial assets. But as $K$ increases the two curves diverge due to the increasing impact of the nonlinearities in $f$, eventually resulting in the appearance of a second minimum when $K$ is just over 3.

When $K$ is between approximately 1 and 4, despite technology $B$ having a larger experience exponent, demand is still too low for it to make enough progress along its experience curve to outweigh its higher variability, so technology $A$ dominates. But as $K$ increases a threshold is crossed ($K \approx 4$), and production suddenly switches to $B$. Thus we observe a demand-driven \emph{unlocking} of technological lock-in, in the sense that if only a small amount of future demand is considered then it is optimal to continue investing in the slower progressing, less uncertain technology, but if market size is large enough then it is optimal to switch to the noisier, faster progressing technology now.

\subsection{Correlated noise}
\label{subsection:correlations}
\begin{figure}
  \centering
  \includegraphics[width=1.0\textwidth]{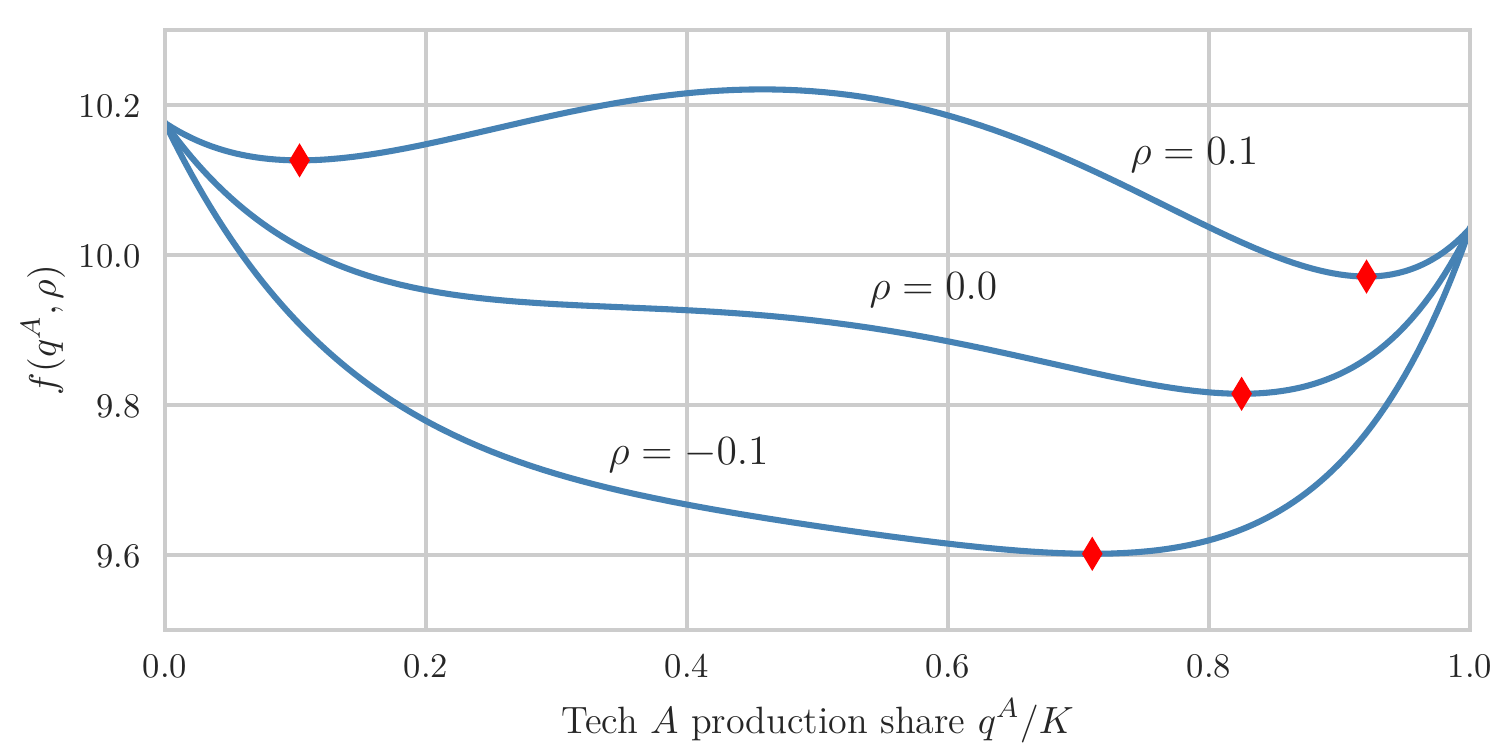}
  \caption{The objective function for three different values of noise correlation $\rho$. Minima are shown in red, risk aversion is fixed at $\lambda = 0.25$, technology $B$ experience exponent is $\alpha^B=0.7$ and all other parameters are as before. The $\rho=0$ line corresponds to the $\alpha^B=0.7$ line in Fig.\ \ref{obj-fn-plot1}. Greater correlation leads to more specialized portfolios and vice versa.}
  \label{obj_fun_plot_alphaB=0.7_correlation}
\end{figure}

In real conditions it is plausible that shocks $\eta^i$ impacting different technologies are not independent. This might be due, for example, to the fact that the same innovations affect both technologies, reducing both costs.
To investigate how the optimization changes in this case we assume that the correlation between $\eta^A$ and $\eta^B$ is $\rho$.
Then the covariance term in $\text{Var} \left( V(q^A) \right)$ is nonzero, so the term
\begin{equation}
2 \lambda q^A q^B c_0^A \left( \frac{z_0^A}{z_0^A + q^A} \right)^{\alpha^A} e^{(\sigma^A)^2 /2} c_0^B \left( \frac{z_0^B}{z_0^B + q^B} \right)^{\alpha^B} e^{(\sigma^B)^2 /2} (e^{\rho \sigma^A \sigma^B} -1)
\end{equation}
must be added to the objective function (Eq.\ (\ref{obj_fn_explicit})).
Fig.\ \ref{obj_fun_plot_alphaB=0.7_correlation} shows the objective function for three different values of $\rho$.
Technology $B$ experience exponent is fixed at $\alpha^B=0.7$ and all other parameters are the same as before, allowing for direct comparison with Fig.\ \ref{obj-fn-plot1}.
Evidently, the effect of increasing the correlation between technology costs is to decrease the benefit of diversification; i.e.\ for fixed risk aversion, increased correlation results in more specialized portfolios, and decreased correlation results in more diversified portfolios.
Indeed, in this example anti-correlation ($\rho=-0.1$) causes $f$ to become convex, and thus have a unique minimum.
(This could otherwise be achieved by increasing $\lambda$.)

Fig.\ \ref{alpha2-lambda-surface} may also be recreated using this version of $f$ (including the correlation term) and the same effect is observed.
This is standard behaviour in a single-period portfolio setting, and does not impact our main findings, so we continue to consider only uncorrelated noise henceforth.

\section{The efficient frontier}
\label{section:efficient_frontier}

Technology portfolios can also be viewed in the efficient frontier framework. This technique is well-known in portfolio theory, and involves plotting each portfolio as a point in expected-return/variance space. We first describe the approach for the restricted Markowitz problem introduced earlier, then show how it applies for technologies (though note that we only consider the no risk-free asset case.)

\subsection{Financial assets}

\begin{figure}
  \centering
  \includegraphics[width=1.0\textwidth]{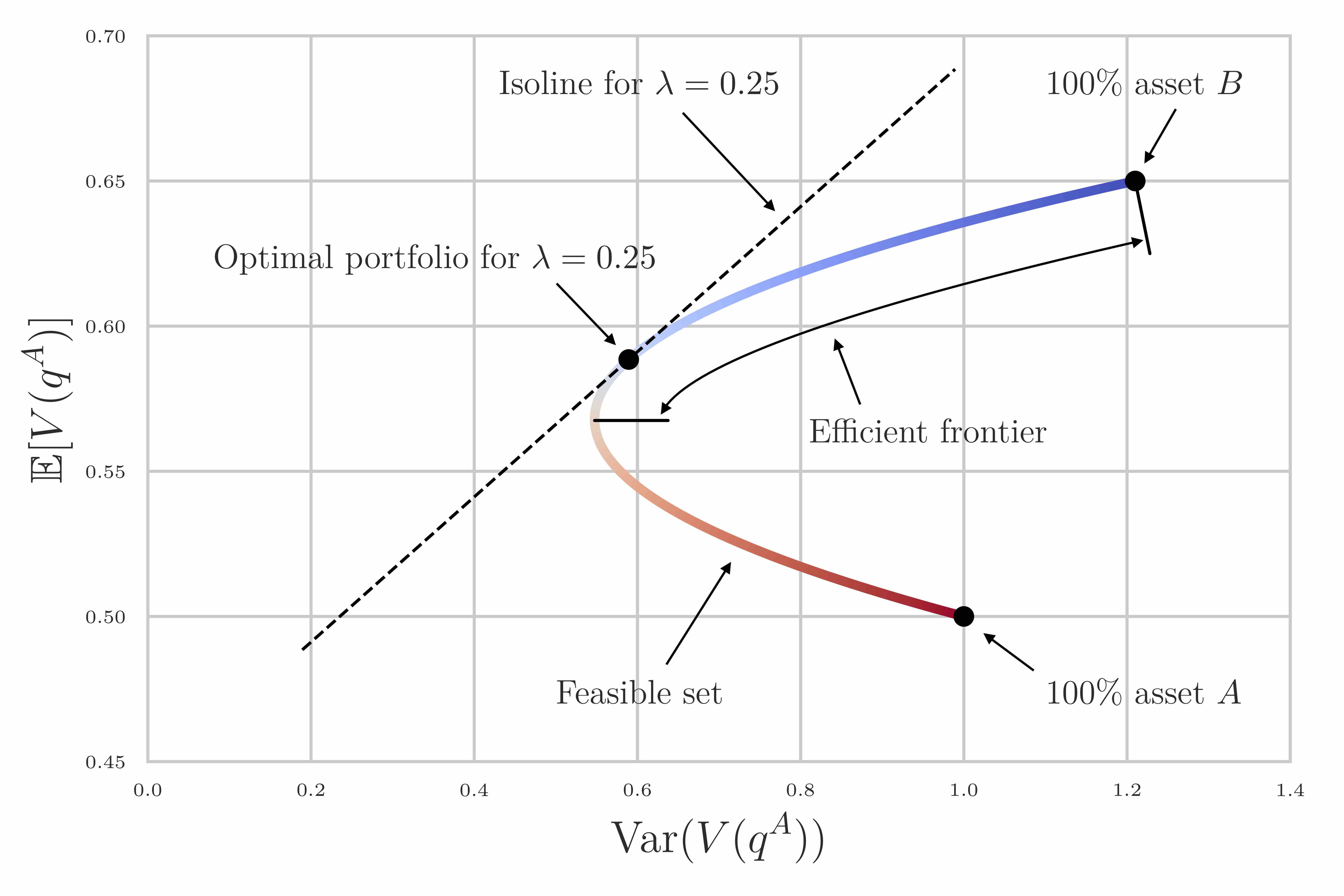}
  \caption{The feasible set of portfolios for the Markowitz system Eq.\ (\ref{markowitz_obj_fn_eq}), with $\mu^A = 0.5$, $\mu^B=0.65$, $s^A=1.0$ and $s^B=1.1$. This is the path of portfolios traced out as the proportion of asset $A$ in the portfolio varies from 0\% (dark blue, $q^A=0$) to 100\% (dark red, $q^A=1$). To demonstrate how risk aversion and optimality are related geometrically, an isoline of $f$ for risk aversion $\lambda=0.25$ is plotted. The black dot at the point of tangency with the feasible set is the unique optimal portfolio for this $\lambda$. The two other black dots represent the two full specialization portfolios.}
  \label{markowitz_frontier}
\end{figure}

In the Markowitz system defined above (Eq.\ (\ref{markowitz_obj_fn_eq})) there are two assets, with known return distributions.
The portfolio weight of asset $A$, $q^A$, is the single free control variable. Each $q^A \in [0,1]$ describes a unique portfolio and as $q^A$ varies from 0 to 1 all feasible portfolios are spanned. For a given value of risk aversion only one of these portfolios is optimal. Each portfolio has a return distribution $V(q^A)$, the expectation and variance of which may be used to plot a single point in expectation-variance space representing the portfolio. This gives the well-known Markowitz diagram: the $x$-axis is the portfolio variance, $\text{Var} \left( V(q^A) \right)$, and the $y$-axis is the expected return, $\mathbb{E} \left[ V(q^A) \right]$. The feasible set of portfolios is the curve traced out on these axes as $q^A$ varies from 0 to 1.
Fig.\ \ref{markowitz_frontier} shows the feasible set for two assets with fixed parameters $\mu^A=0.5$, $\mu^B = 0.65$, $s^A=1.0$ and $s^B=1.1$ (\textit{cf.}\ Section \ref{subsection:markowitz}).
This horizontal parabola is known as the Markowitz bullet. The colour scheme is the same as before (\textit{cf.}\ Fig.\ \ref{markowitz_plot}) so dark red corresponds to 100\% asset $A$ and dark blue to 100\% asset $B$.

From the definition of $f$ (Eq.\ (\ref{markowitz_obj_fn_eq})) we have $\mathbb{E} \left[ V \right] = f(V) + \lambda \text{Var} \left( V \right)$, so on these axes the isolines of $f$ (i.e.\ level sets of $f$), for any fixed value of risk aversion, are just the straight lines of gradient $\lambda$.
The value of $f$ for each portfolio lying on a given isoline (i.e.\ the points of intersection with the feasible set) is given by the $y$-axis intercept. Then since we want to \textit{maximise} $f$ in this problem, the optimal portfolio for this $\lambda$ is the unique point of intersection of the feasible set and the isoline of gradient $\lambda$ with highest $y$-axis intercept.
The \textit{efficient frontier} is defined as the set of all portfolios which are optimal for some value of $\lambda$. Therefore, since $\lambda \in [0, \infty)$, the efficient frontier here is the segment of the feasible set furthest into the upper-left-most quadrant of the diagram. These elements are all shown on Fig.\ \ref{markowitz_frontier}.

\subsection{Technologies}
\label{markowitz_plot_technologies}
\begin{figure}
  \centering
  \includegraphics[width=1.0\textwidth]{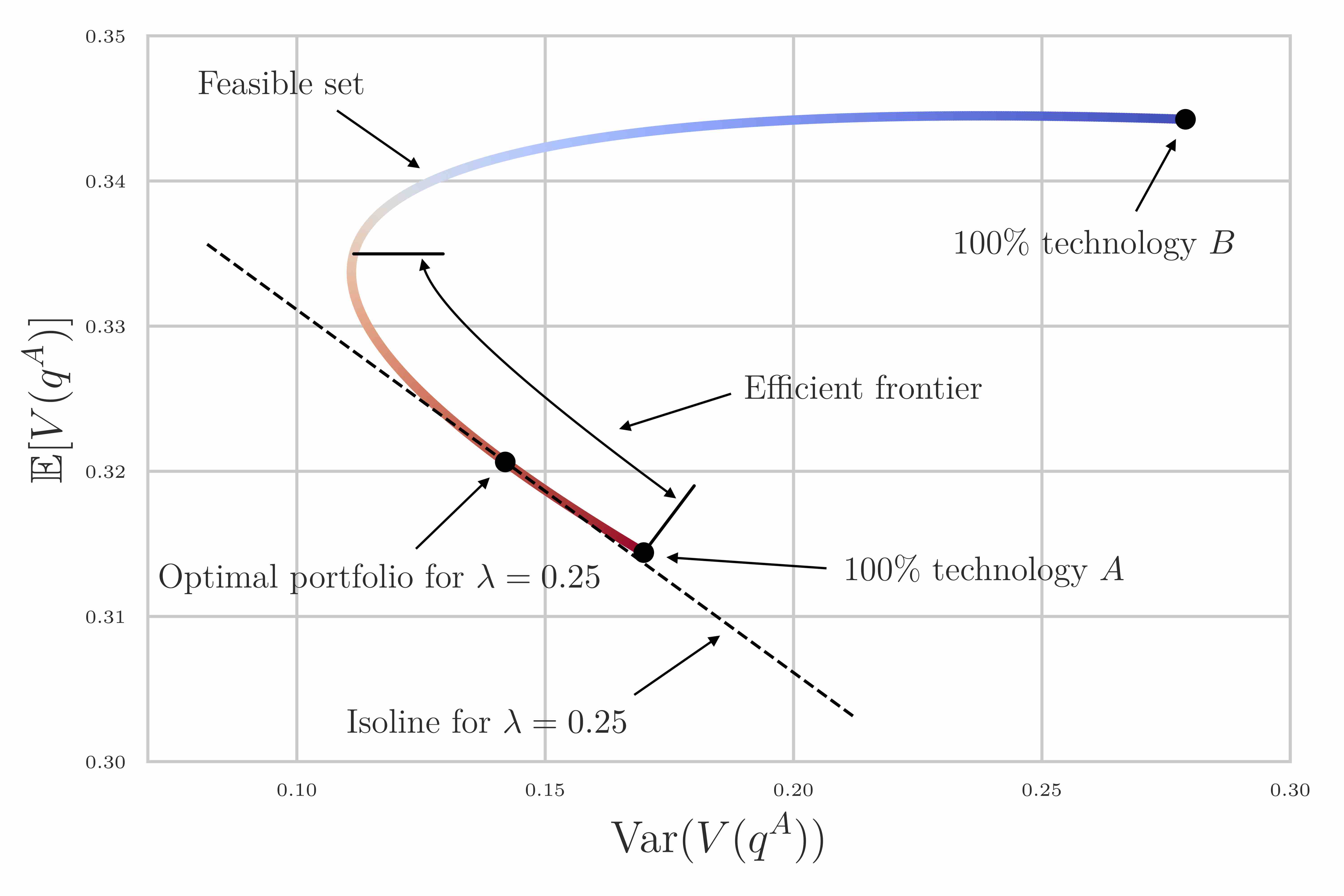}
  \caption{The feasible set of portfolios for two technologies in a low-learning regime.
This is the path of portfolios traced out as the proportion of technology $A$ production in the portfolio varies from 0\% (dark blue, $q^A=0$) to 100\% (dark red, $q^A=K$).
The technologies here have $\alpha^A = 0.5$, $\alpha^B=0.65$, and other parameters are those shown in Table \ref{alpha2-lambda-surface-param-values}, except demand, which is set to $K=0.1$.
This severely limits the potential for learning, so the problem is nearly-Markowitz and hence the feasible set is almost parabolic.
Isolines of $f$ now slope downward and the efficient frontier is the lower-left-most portion of the feasible set.
An isoline corresponding to risk aversion $\lambda=0.25$ is plotted. The black dot at the point of tangency with the feasible set is the unique optimal portfolio for this $\lambda$.}
  \label{portfolio_frontier1}
\end{figure}

\begin{figure}
  \centering
  \includegraphics[width=1.0\textwidth]{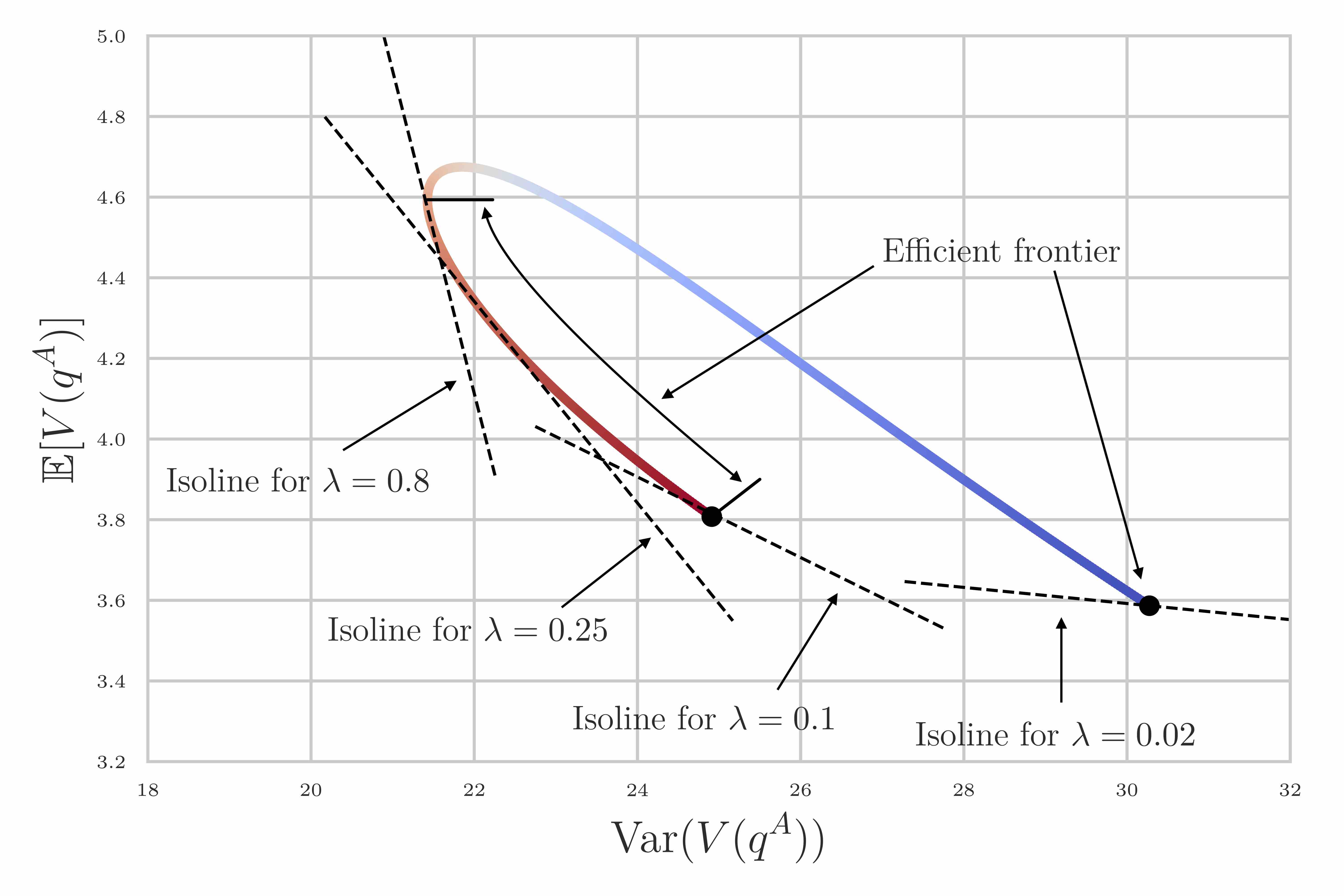}
  \caption{The feasible set of portfolios for technologies analogous to Fig.\ \ref{portfolio_frontier1}, but with demand now set at the higher value $K=2$.
The technologies are no longer in a low-learning regime, and nonlinear feedback causes the feasible set to be stretched and tilted.
Isolines for four values of risk aversion are plotted, showing how the transformed geometry of the feasible set causes the efficient frontier to now consist of two disconnected components.
There is a critical value of risk aversion ($0.02 < \lambda_{\textit{switch}} < 0.1$) at which two optimal portfolios exist simultaneously ($100\%$ tech $A$ and $100\%$ tech $B$).
This is how instantaneous optimum switching is manifested in expectation-variance space.
}
  \label{portfolio_frontier2}
\end{figure}

In contrast to the Markowitz model, in the technologies model we want to minimize both the variance \textit{and} the expected cost, so the sign of the variance part of the objective function is reversed.
The isolines of $f$ are therefore now the \textit{downward}-sloping straight lines, of gradient $-\lambda \in (-\infty, 0]$. And since lower $f$ is now better, an optimal portfolio is a point of intersection of the feasible set with the isoline of \textit{lowest} $y$-axis intercept. The efficient frontier therefore consists of the part of the feasible set furthest into the lower-left-most quadrant of the diagram.
We demonstrate these differences between the Markowitz and technology models by first plotting the expectation-variance diagram for two technologies in a low-learning regime, and then for the same two technologies in a high-learning regime.
These are shown in Figs.\ \ref{portfolio_frontier1} and \ref{portfolio_frontier2}, and are analogous to Fig.\ \ref{markowitz_frontier}.

In Fig.\ \ref{portfolio_frontier1}, technology $B$ has experience exponent $\alpha^B = 0.65$ and other parameters are those shown in Table \ref{alpha2-lambda-surface-param-values} as before, except for demand, which is set to $K=0.1$. By restricting production in this way, learning effects are very small, so we are in a nearly-Markowitz regime, and hence the feasible set is almost parabolic. The figure makes clear how the efficient frontier and isolines differ in the technologies problem, as compared to the Markowitz problem (Fig.\ \ref{markowitz_frontier}).
In Fig.\ \ref{portfolio_frontier2} all parameters are identical to Fig.\ \ref{portfolio_frontier1}, except demand, which is now returned to the value in Table \ref{alpha2-lambda-surface-param-values}, $K=2$, so that the technologies are no longer in a low-learning regime.

As these plots show, there are two significant differences between how technologies and financial assets appear in this framework. First, for technologies the feasible set is tilted and stretched. This is because the objective function is highly nonlinear in portfolio weights, not just quadratic. Second, as a direct consequence of this, the efficient frontier may now be split into two disconnected components. This is the case in Fig.\ \ref{portfolio_frontier2}, where the efficient frontier consists of both the long red segment on the left (mainly technology $A$) \textit{and} the isolated end-point on the right ($100\%$ technology $B$). This splitting of the efficient frontier is how instantaneous optimum switching is manifested in expectation-variance space: as risk aversion goes from $\lambda = 0$ to $\infty$ the optimal portfolio traverses the efficient frontier from one end to the other, jumping from one component to the other at the critical value $\lambda = \lambda_{\textit{switch}}$.
At the point of the discontinuity $f$ has two distinct minima of equal value, and there are two optimal portfolios, both of which lie on the same isoline, of gradient $-\lambda_{\textit{switch}}$.
In this case these are the 100\% $A$ and 100\% $B$ portfolios.

Note that since $\alpha^B=0.65$ in Fig.\ \ref{portfolio_frontier2}, the efficient frontier shown corresponds to the $\alpha^B=0.65$ section through the surface in Fig.\ \ref{alpha2-lambda-surface}.
Hence the change in the optimal portfolio as risk aversion varies can be traced out equivalently on both diagrams.
The closed-form expression for $\lambda_{\textit{switch}}$ is given in Appendix \ref{subsubsection:discontinuity}.

Viewing the problem in the expectation-variance framework shows how optimal technology portfolios of equal value can coexist simultaneously (unlike in financial portfolios). Similar value portfolios may have either large expectation and small variance, or vice versa, or some combination in between. And since the feasible set is no longer parabolic (due to Wright's law nonlinearities), there may be many very different portfolios lying near the optimal isoline. For example, in Fig.\ \ref{portfolio_frontier2} all portfolios with around 60-90\% technology $A$ (red) lie very near to the optimal $\lambda=0.25$ isoline, because the feasible set has very low curvature here.

This is suggestive of the behaviour we would expect to see in a multi-technology model. The increasing returns dynamic allows many different ways of generating portfolios of similar value, using different combinations of the various technologies' expectations and variances. This would result in a highly non-convex optimization problem with many local minima.

\subsection{Effect of demand on the efficient frontier}
\label{demand_efficient_frontier}
\begin{figure}
  \centering
  \includegraphics[width=1.0\textwidth]{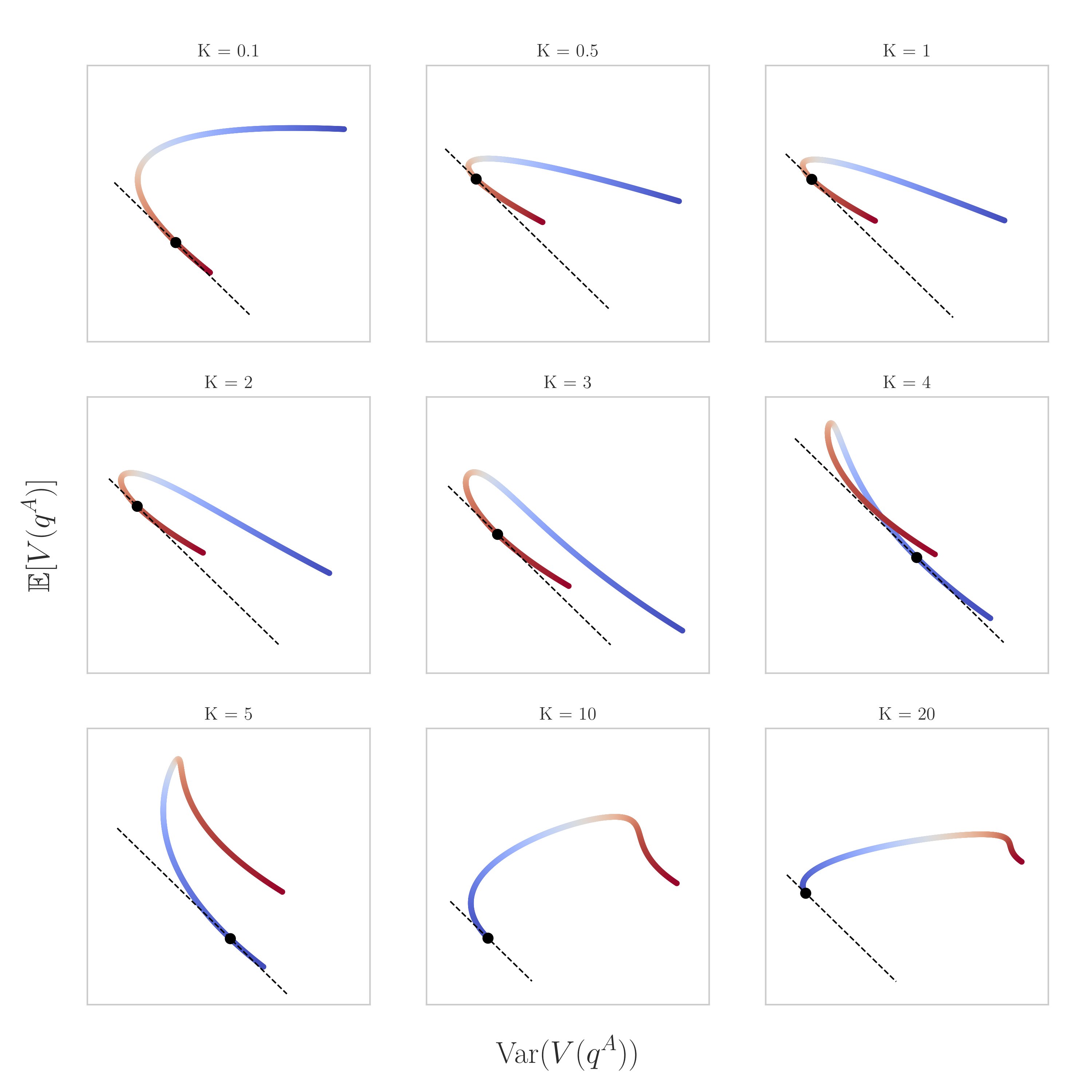}
  \caption{Feasible portfolios for $\alpha^A=0.5$, $\alpha^B=0.65$ and varying $K$,  with other parameters fixed as before (Table \ref{alpha2-lambda-surface-param-values}). Again, dark blue corresponds to 0\% technology $A$ ($q^A=0$) and dark red to 100\% ($q^A=K$). The dashed lines are $\lambda=0.25$ isolines of $f$, and the black dots are the
  optimal portfolio for this risk aversion. The axes scales have been omitted for clarity (they are different for each plot). These plots show how the problem transitions from a Markowitz-like, low-learning regime when $K$ is small, to a highly nonlinear high-learning regime when $K$ is large.}
  \label{portfolio_frontier_multi}
\end{figure}

Fig.\ \ref{portfolio_frontier_multi} shows in more detail the effect of total demand on the feasible set and optimal portfolios.
The technologies are fixed, and are the same as in Figs.\ \ref{portfolio_frontier1} and \ref{portfolio_frontier2}. $K$ is the only parameter which varies. Although the scales on the axes differ in each plot, the ratio between them is constant. Indeed, the dashed lines all have gradient $\lambda=-0.25$; they are the isolines corresponding to the optimal portfolio for $\lambda=0.25$ in each case.

When $K$ is tiny there is very little potential for learning so the technologies behave in an ``asset-like'' way, and the feasible set is almost a parabolic Markowitz bullet. Here technology $A$ (red) dominates the optimal portfolio for all values of risk aversion. As $K$ increases the potential for learning increases, so the nonlinearities in $f$ start to have an impact and the feasible set starts to become distorted. At around $K=1$ the blue arm drops below the red arm, splitting the efficient frontier in two and indicating the presence of two equal value portfolios for the first time. Here, for $\lambda \leq \lambda_{\textit{switch}}$ 100\% technology $B$ (blue) is optimal, but for $\lambda \geq \lambda_{\textit{switch}}$ technology $A$ (red) is dominant. At around $K=4$ the two arms cross and large portions of the feasible set are very close to the $\lambda=0.25$ isoline. Hence there are many different nearly-optimal portfolios in this case. After this point the two arms of the feasible set cross over completely so that technology $B$ is dominant for all values of risk aversion. As $K$ gets very large ($> 10$) the potential for learning is so great that near full specialization in the fastest-learning technology ($B$) is optimal for all levels of risk aversion.
Fig.\ \ref{portfolio_frontier_multi} may be related back to Fig.\ \ref{cusp_slice_K} (though note that distance increments \emph{along} the feasible set do not correspond linearly to increments in $q^A/K$).

Clearly this analysis relies on the assumption that noise is independent of total production $K$, so that more production, and hence learning, can take place without affecting the size of the shocks.

\section{Asymmetrical technologies and escaping lock-in}
\label{section:lockin}
The paper so far has focused on the case of two almost symmetrical technologies, studying the behaviour of the optimal portfolio as one parameter is changed, while all others are held constant.
We now consider a more realistic and interesting example in which an established technology is challenged by a newcomer\footnote{The limiting case of this is when one technology is a `safe' technology, with constant cost ($\alpha^A = \sigma^A =0$). Then, in the nearly-Markowitz regime ($K \ll z_0^B$), the objective function reduces to
\begin{equation}
f \approx c_0^A \left( K- q^B \right) + c_0^B e^{ (\sigma^B)^2 / 2} q^B + \lambda \left( c_0^B \right)^2 e^{ (\sigma^B)^2} \left( e^{(\sigma^B)^2} -1 \right) (q^B)^2
\label{markowitz_approx_safe}
\end{equation}
(\emph{cf}.\ Eq.\ (\ref{markowitz_approx})). This is a convex parabola with minimum
\begin{equation}
q_{*}^B = \frac{c_0^A - c_0^B e^{(\sigma^B)^2/2}}{2 \lambda \left( c_0^B \right)^2 e^{(\sigma^B)^2} \left( e^{(\sigma^B)^2}-1 \right) },
\label{safe_minimum}
\end{equation}
and the condition for the optimal solution to be diversified between the safe and the new technology is $0 < q_{*}^B < K$. This is analogous to the Markowitz portfolio problem with a safe asset and a risky asset. Outside of this low-learning regime though, the safe technology simplification does not result in increased analytical tractability, so the numerical solution approach is still applicable.
}.
Suppose the setting is one where a cheap, mature, slow-learning technology $A$ is challenged by a costly, young, fast-learning technology $B$. Table \ref{asym-tech-param-values} shows the parameter values used here.
\begin{table}[H]
\centering
\begin{tabular}{|clcc|}
\hline
Symbol & Description & Tech $A$ & Tech $B$\\
\hline
$z_0$ & Technology maturity & 100 & 1\\
$c_0$ & Initial cost & 1& 2\\
$\alpha$ & Experience exponent & 0.15 & 0.2 \\
$\sigma$ & Technology volatility & 0.1 & 0.1\\
\hline
$K$ & Demand &\multicolumn{2}{c|}{[0-100]} \\
$\lambda$ &Risk aversion&\multicolumn{2}{c|}{[0-1.2]} \\
\hline
\end{tabular}
\caption{Parameter values for the case where a young expensive technology competes with an old cheap technology.}
\label{asym-tech-param-values}
\end{table}

\begin{figure}
  \centering
  \includegraphics[width=1.0\textwidth]{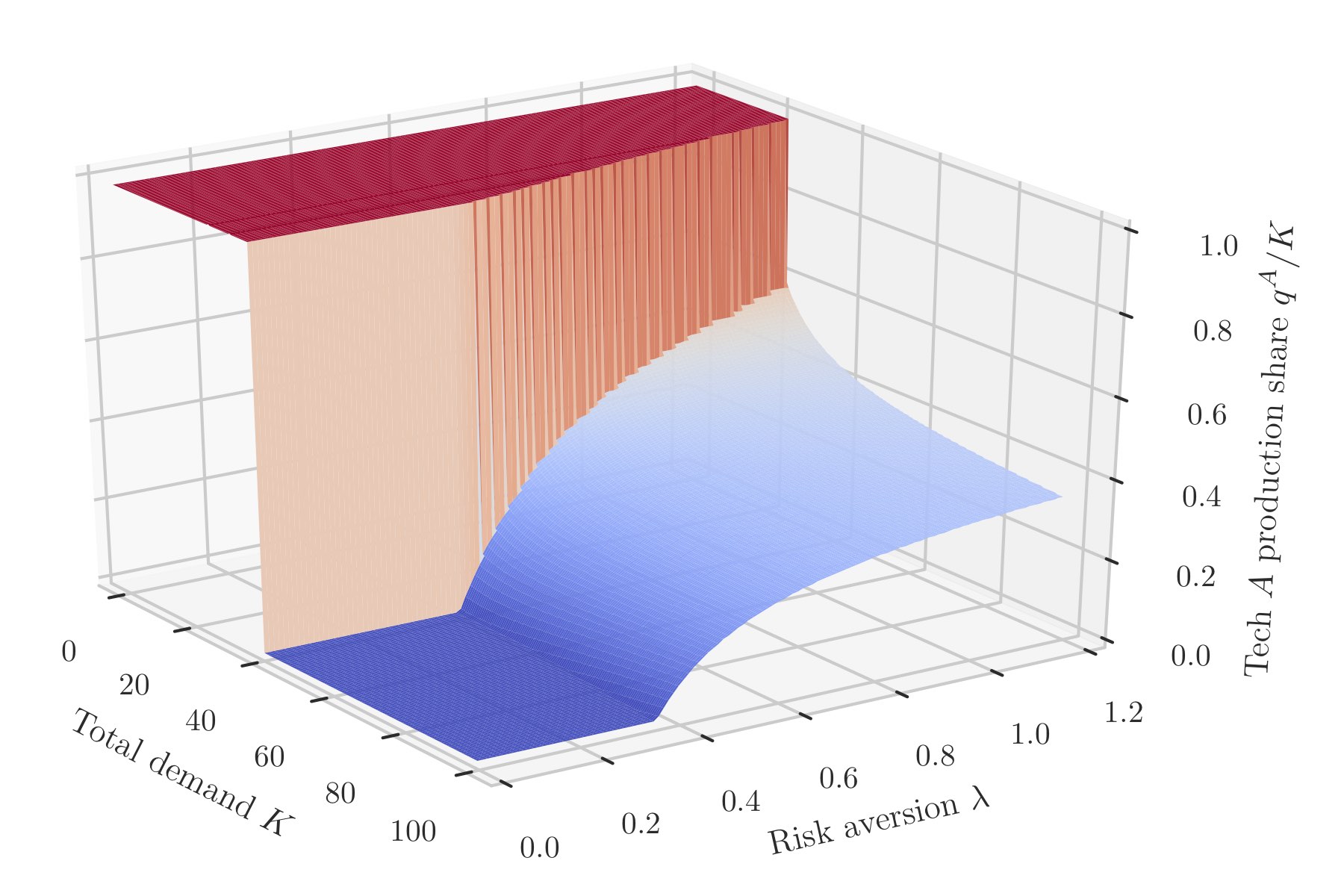}
  \caption{Optimal share of production of incumbent technology $A$ (red) when in competition with challenger technology $B$ (blue), for varying demand $K$ and risk aversion $\lambda$. Parameter values are shown in Table \ref{asym-tech-param-values}. Technology $A$ has such a strong initial cost advantage that when demand is low (and hence the potential for learning is low), it is optimal to specialize fully in $A$, for all values of risk aversion shown. As demand increases, so does the potential for learning, and in order exploit the faster-learning challenger the global optimum switches to a new local minimum, in which technology $B$ dominates.}
  \label{lambda_K_surface}
\end{figure}

Repeating the demand-driven lock-in analysis of Section \ref{demand_section}, Fig.\ \ref{lambda_K_surface} shows the optimal portfolio surface over total demand and risk aversion axes.
Technology $A$'s initial cost advantage is so strong that a demand of at least 20 times the initial cumulative production of technology $B$ is required to prevent full specialization in $A$, even for high risk aversion. We observe the familiar optimum switching as demand increases, demonstrating again how important the role of anticipated future demand is in determining the optimal production mix. In this case, the $K$-$\lambda$ parameter space separates into three qualitatively different regimes: \textit{i}) for low $K$ 100\% technology $A$ is optimal, \textit{ii}) for high $K$ but low risk aversion 100\% technology $B$ is optimal, and \textit{iii}) for high $K$ and at least moderate risk aversion, the optimal proportion of technology $A$ is around 0-40\%.
Results like this could be very useful in applications, where often the key challenge is simply reducing the dimension of the decision space.

For large $K$ there is a qualitative change in the surface at $\lambda \approx 0.4$.
Here the global optimum of the objective function moves from the boundary to the interior of the optimization range (the value of $\lambda$ at which this transition occurs may be found analytically as shown in Appendix \ref{subsection:onset_of_diversification}).
This is similar to the situation in Fig.\ \ref{obj-fn-plot1}, where as $\alpha^B$ increases from $0.71$ to $0.72$ up to, say $0.9$ (not shown), the global minimum moves towards the boundary, then sticks on it (and in fact the function minimum over $\mathbb{R}$ continues to move further outside the optimization range, but this is not a valid solution).

For comparison, Appendix \ref{appendix:demand_risk_surface_similar_techs} shows the same demand-risk surface for the case of the two almost identical technologies of Section \ref{section:t1}.

To summarise the behaviour of the model, note the direction in which each of the parameters would need to move in order to escape lock-in to an incumbent technology. All things being equal, avoiding lock-in to technology $A$ would require: decreased $B$ maturity, decreased $B$ initial cost, increased $B$ experience exponent, decreased $B$ volatility, increased demand $K$, or increased risk aversion $\lambda$.

\section{Two-period model}
\label{section:two_period_model}
Having analysed the model in the simplest, single-period case, we now consider the extension to two periods.
This allows us to introduce discounting and investigate its effect on optimal production timing.
This is especially relevant for experience curves since they exemplify the concept of investing effort now to unlock future benefits, so the relative value of present and future benefits is critical.
Moving to a multi-period setting allows us to investigate the conditions under which we should plan to invest in a technology now, or in future (or neither), given current knowledge about the present state of technologies and their likely development under various investment scenarios.
We briefly discuss some features and limitations of our approach.

The extension we consider is static, in the sense that the model schedules production now, \emph{for all future periods}, in the optimal manner as defined by the objective function.
This static analysis relies on current estimates of technology parameters, which are in turn based on empirical data about the technologies in question.
But as time progresses we observe realisations of the noise, thereby gaining new information about technology costs and parameters.
Therefore performing the same optimization procedure in future may yield different results, and we are faced with the possibility of time-inconsistency.
This is a limitation of the static model\footnote{The dynamic mean-variance portfolio problem is difficult even in the case of standard financial returns, see e.g.\ \citet{garleanu2013_dynamictrading}. In our case the complexities are increased by the nonlinearities and time dependencies in costs. One way to approach the problem would be to discretize the decision space and the random event space, then apply backward induction to the resulting “scenario tree”. See, e.g., \citet{Edirisinghe2007} for an application of this method to the standard mean-variance portfolio problem. However, even for rough discretization the resulting tree would grow very quickly. Hence, we do not pursue this analysis here, but leave it for future research.}.
In support of the static approach however, note that portfolio adjustment costs could be very high for technologies, which may result in high levels of commitment for future periods anyway.

In this model, technological progress only occurs via the stochastic experience curve mechanism, there is no exogenous progress trend.
This is due to our choice of zero mean noise, $\eta  \sim \mathcal{N}(0,\sigma^2)$, which we use because it is close to the model tested empirically by \citet{lafond2018}.
By using normal noise with nonzero mean instead it is possible to model an exogenous progress trend.
For $\eta  \sim \mathcal{N}(\mu,\sigma^2)$, the standard lognormal distribution has expectation $e^{ \mu + \sigma^2 / 2}$, so if $\mu < - \sigma^2 / 2$ then the expected cost of a technology can in fact decrease between periods under zero production (\emph{cf}.\ Eq.\ (\ref{cost_expectation})).
This is especially important in the multi-period setting (although it also applies in the single-period case).
If a technology is initially very expensive, and $\mu$ is very negative, then waiting for the technology to improve may indeed be a viable strategy.
However, while waiting for this improvement in the expectation, the cost variance may become less favourable (relative to that of the other technology), counteracting any benefit in the objective function.
Generally, increasing experience in one technology comes at the expense of the other, and if it is optimal to ``delay production'' in one technology then this is due only to the complex interplay of all model parameters in the objective function, not because of some simple background improvement in a technology.
Here though we continue to use zero mean noise.

For simplicity, as in the one-period case, we use technology production, not investment, as a proxy for experience.
In the single period case the distinction is not significant, but when considering multiple periods it is, due to capital depreciation.
Thus our multi-period framework does \emph{not} model investment strategies, only production strategies.
A proper treatment of investment would need to model the relationship between production and investment, which would require at least one extra parameter to represent depreciation.
We prefer not to complicate the model further, and instead just consider production, while highlighting this discrepancy.

\subsection{Optimization}
\label{section:2period_optimization}

We use the same first-difference Wright's law model as before (Eq.\ (\ref{log_first_diff_eqn})), with subscripts now denoting distinct periods, and separate noise shocks impacting each period:
\begin{equation}
\label{log_first_diff_eqn2}
\log(c_{t}^i)- \log (c_{t-1}^i) = -\alpha^i \big[ \log(z_{t}^i) - \log (z_{t-1}^i) \big] +\eta_{t}^i \ , \qquad i=A,B, \ t=1,2.
\end{equation}
Production of each technology accumulates in the obvious way: $z_t^i = z_0^i +  \sum_{k=1}^t q_k^i$.
We retain the assumption that each technology has its own specific noise distribution, $\eta^i  \sim \mathcal{N}(0,(\sigma^i)^2)$, 
and now assume that in each period there is a new draw from this distribution, $\eta_t^i$.
For simplicity we assume shocks are uncorrelated over both time and technologies (though clearly correlated noise has the potential to play an important role in a multi-period setting).
Average unit costs in the second period are then given by:
\begin{equation}
c_2^i  \ = \ c_1^i \left( \frac{z_1}{z_2} \right)^{\alpha^i} e^{\eta_2^i} \ = \ c_0^i \left( \frac{z_0}{z_2} \right)^{\alpha^i} e^{\eta_1^i + \eta_2^i}.
\label{cost2}
\end{equation}
This makes it clear how successive shocks impact cost in the two-period setting: from period to period costs fall according to Wright's law, but the periodic shocks accumulate, potentially driving the cost far from the deterministic experience curve trend. 
In contrast to standard experience curve implementations, this has the advantage that costs are not guaranteed to fall in the long run due to experience effects, and instead admits the possibility that uncertain exogenous events may dominate.

While it is possible to allow any of the parameters to vary over the periods in this setup (e.g.\ experience exponents, noise distributions), here we keep them all fixed, again for simplicity and clarity.
Total production is set to $K$ \emph{per period}, so there are now two production constraints: $q_t^A+q_t^B  =K$ for $t=1,2$.
The above choices are made because the model gets unwieldy very quickly, so it is preferable to use the simplest possible formulation that captures the key features of interest.

We implement exponential discounting with discount rate $r$ and consider the present discounted cost of the total system:
\begin{equation}
V(q_1^A, q_2^A) = \sum_{t=1,2} \sum_{i=A,B} e^{-r(t-1)} c_t^i q_t^i.
\label{total_cost2}
\end{equation}
In contrast to the one-period total cost (Eq.\ (\ref{total_cost})), this is now a function of production in both periods.
We use the same mean-variance objective function as before, details of which are given in Appendix \ref{appendix:2period_obj_fn}\footnote{As mentioned earlier, the choice of objective function in a multi-period setting is more subtle and important than for a single period; we leave this consideration for future research.}. The optimization problem is then
\begin{eqnarray}
\label{obj_fn_simplest_form2}
&\underset{q_1^A, q_2^A}{\text{minimize:}} & f(q_1^A, q_2^A) = \mathbb{E} \left[ V(q_1^A, q_2^A) \right] + \lambda \text{Var} \left( V(q_1^A, q_2^A) \right) \\[0.3cm]
&\text{subject to:}& q_1^A \in [0, K], \ q_2^A \in [0, K]. \nonumber
\end{eqnarray}

\subsection{Results}
\label{section:2period_results}
To show how discount rate $r$ and risk aversion $\lambda$ interact to influence optimal production over the two periods we fix all other technology parameters and production constraints, then plot the objective function for a range of $r$-$\lambda$ pairs.
Since $f$ is now a function of the two control variables, $q_1^A$ and $q_2^A$, we plot its values as a contour-/heat-map over these axes.

\begin{figure}
  \centering
  \includegraphics[width=1.0\textwidth]{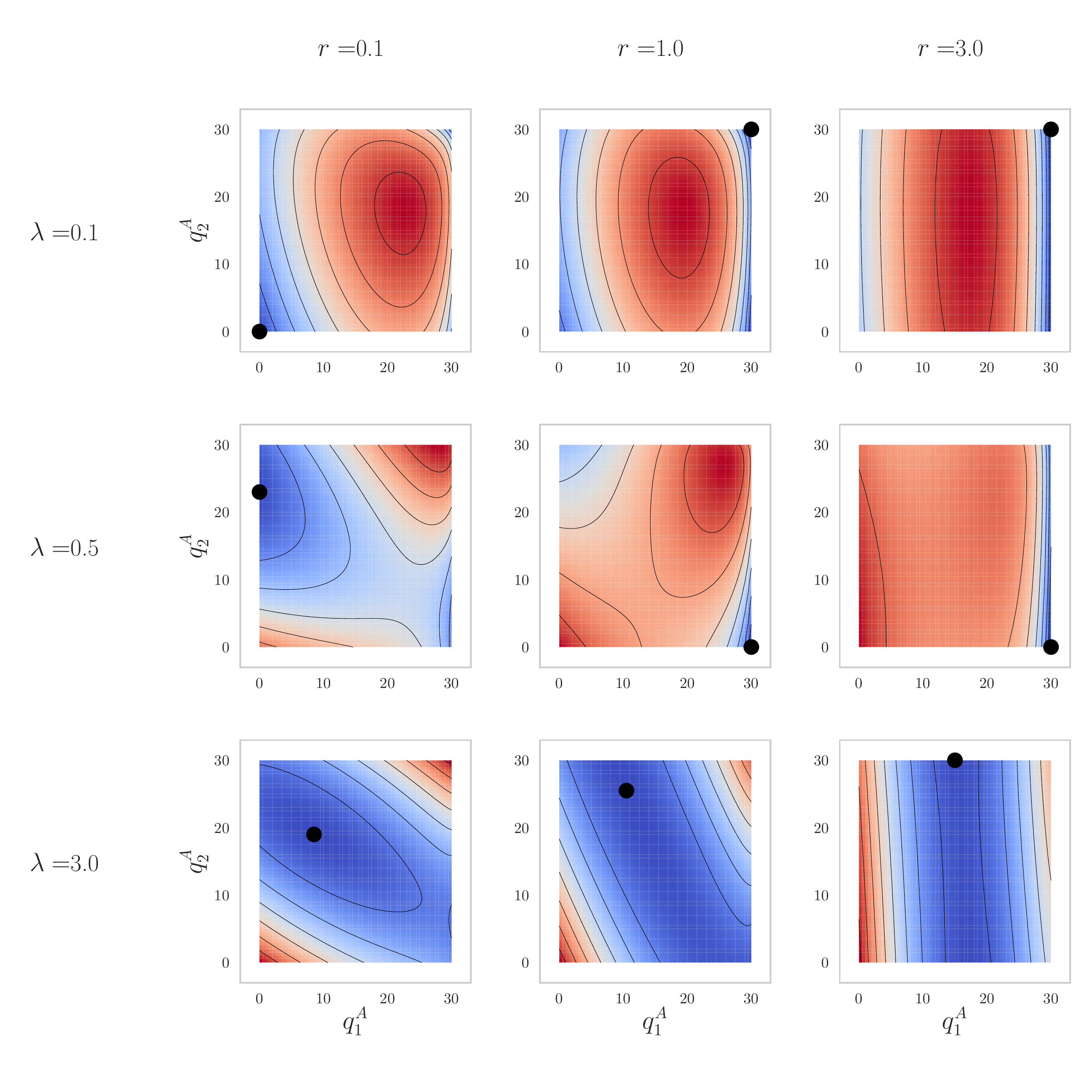}
  \caption{Heat-maps showing the value of the two-period objective function for a range of risk aversion and discount values, as a function of first and second period production in technology $A$ ($q_1^A$, $q_2^A$). Blue corresponds to lower values of $f$, red to higher values, and the global optimum is shown as a black dot. The technologies used here are the asymmetrical technologies of Section \ref{section:lockin}: $A$ is a cheaper, mature, incumbent technology and $B$ is a young, fast-learning challenger.}
  \label{two_period_contour_plots}
\end{figure}

Fig.\ \ref{two_period_contour_plots} shows the results for nine pairs of low, medium and high discount rate and risk aversion: $r=0.1,1.0,3.0, \ \lambda = 0.1,0.5,3.0$.\footnote{These values are picked for illustrative purposes only, in order to display various features clearly. However, to justify these discount rates, it is useful to think about the length of one period being about 30-40 years. This could be motivated by long horizons in the energy sector; for instance, thermal power plants are designed for a life of 30 to 40 years.}
The two technologies used here are the same as in the asymmetrical case of Section \ref{section:lockin} (Table \ref{asym-tech-param-values}), and total production $K$ is set to 30 units per period, as this allows results to be related directly back to the one-period case in Fig.\ \ref{lambda_K_surface}. (Production is shown here in units instead of percentages for easier comparison.)
Darker blue corresponds to lower values of $f$ and darker red to higher values, though the scale is different for each plot.
The global minimum, again obtained by brute force optimization, is shown as a black dot (though this sometimes forms part of a valley of nearly identical values).

\subsection{Discussion}
\label{section:2period_discussion}
We observe the same kind of phenomena as in the one-period case: highly specialized solutions for low risk aversion (due to deterministic learning feedback), greater diversification for high risk aversion (due to the counteracting effects of variance terms in $f$), multiple optima of the objective function, and discontinuities in the global optimum as parameters vary.

First note that as the discount rate increases the two-period problem reduces to the one-period problem.
Plots in the $r=3.0$ column are nearly uniform in the $q_2^A$ direction, since second period costs are discounted so heavily that the choice of second period production makes little difference to $f$.
The global optima here correspond to the global optima shown at the relevant points on Fig.\ \ref{lambda_K_surface}.

To observe the presence of multiple optima and instantaneous switching of the global minimum, consider the two left-most plots of the top and middle rows of Fig.\ \ref{two_period_contour_plots}.
In the top row ($\lambda=0.1$), the two plots are formed of concentric oval isolines, with a maximum of $f$ at their centres.
All four corners are local minima, reflecting the fact that strong feedback in the deterministic experience curve setting discourages any diversification at all.
As $r$ goes from $0.1$ to $1.0$, the global optimum switches from $(0,0)$ to $(30,30)$ (similarly to Fig.\ \ref{obj-fn-plot1}).
This is because when second period costs are only weakly discounted, technology $B$'s higher experience exponent has the potential to generate sufficiently low expected second period cost that it is worth concentrating production in $B$ in the first period, then reaping the rewards in the second.
But when second period costs are moderately discounted, this second period `reward' has lower overall value in $f$, so it is not worth producing any technology $B$ at all in either period. (This is just standard deterministic experience curve behaviour with discounting.)

In the middle row ($\lambda=0.5$) we again observe optimum switching behaviour, but the situation is more nuanced.
$f$ is now a saddle, with opposite corners forming distinct minima and maxima as the different expectation and variance components of $f$ contribute and interact in complex ways.
In comparison with the top row though, the central areas of these plots are generally less unattractive, since risk aversion allows more balanced portfolios.
As $r$ varies from $0.1$ to $1.0$, the global optimum switches from around $(0,23)$ to $(30,0)$.
When discounting is very weak the optimal strategy is again to concentrate production in technology $B$ initially in order to bring the cost down, but then to diversify slightly in period two in order to lower the variance.
When discounting is moderate though, the expected cost reduction in technology $B$ is not valuable enough to override the initial cost advantage of $A$, so first period production is concentrated in $A$.
Then however, production switches to being entirely concentrated in technology $B$ in the second period.
This is likely due to the fact that the covariance term in $f$ (see Eq.\ (\ref{full_two_period_variance})) contains the product $q_1^A q_2^A$, so there is a benefit to setting one of these terms equal to zero, and in this particular case the effect is large enough to produce the bang-bang solution shown.
However, since all 10 components of $f$ (see Appendix \ref{appendix:2period_obj_fn}) are being traded off against each other in this moderate risk aversion/ moderate discounting regime, the relative size of any specific effect is not clear (without further analysis).

Finally, in the bottom row ($\lambda=3.0$), risk aversion is strong and hence diversification is optimal. Whole regions of the solution space generate identical or similar values of $f$. This is because, as in the one-period case, expected progress is being traded off against the certainty in this progress, and the same objective function value may be achieved by many different combinations of first and second period production.

As in the one-period problem, due to the functional form of the objective function, the instantaneous optimum-switching behaviour demonstrated here with the $r$ and $\lambda$ parameters can also occur when any one of the underlying parameters is varied continuously.

\subsection{Effects of discounting and risk aversion in scenario comparison}
\label{section:2period_scenario_comparison}
While optimization over the entire solution space is essential for understanding the basic properties of the model, there are situations in which this is either undesirable or impossible, and it is better simply to compare the performance of a restricted set of portfolios. We refer to this as scenario comparison (as this captures better the idea of comparing worlds characterised by different technology mixes). This can be useful either if there are additional system constraints, or if the solution space is very large, for the following reasons.

First, in practise technologies are not perfect substitutes, so although they may be considered substitutes in terms of some primary characteristic (i.e.\ ``production''), other, secondary characteristics may require portfolios to be constrained in additional ways.
For example, solutions that vary wildly from period to period may be impractical in many real world situations.
Or as another example, consider a portfolio of energy technologies on an electric power grid.
While they may be considered substitutes in terms of, say, total annual energy production, many other engineering and physical constraints must be met in order to ensure grid stability\footnote{E.g.\ if large shares of intermittent renewable technologies are present then large quantities of energy storage or other backup technologies may also be required, depending on daily and seasonal demand patterns.}.
This would place extra restrictions on such technology portfolios; these constraints could of course be modelled explicitly and a more complicated objective function constructed, but this is outside the scope (and spirit) of the work here.

Second, consider extending the model to include more technologies and/or periods.
As the number of control variables increases, brute force optimization quickly becomes computationally intractable, and since the problem is non-convex (due to the experience curve feedback), local optimization methods are not guaranteed to find the global optimum. It may therefore be necessary to use heuristic arguments to specify a restricted set of available portfolios, and compare these directly.

Now, since portfolios are fixed in the scenario comparison setting, this allows for an alternative perspective on the effects of discounting and risk aversion.
We demonstrate this with an example, using the same two-period model and asymmetrical technologies as earlier in this section.

\begin{figure}
  \centering
  \includegraphics[width=1.0\textwidth]{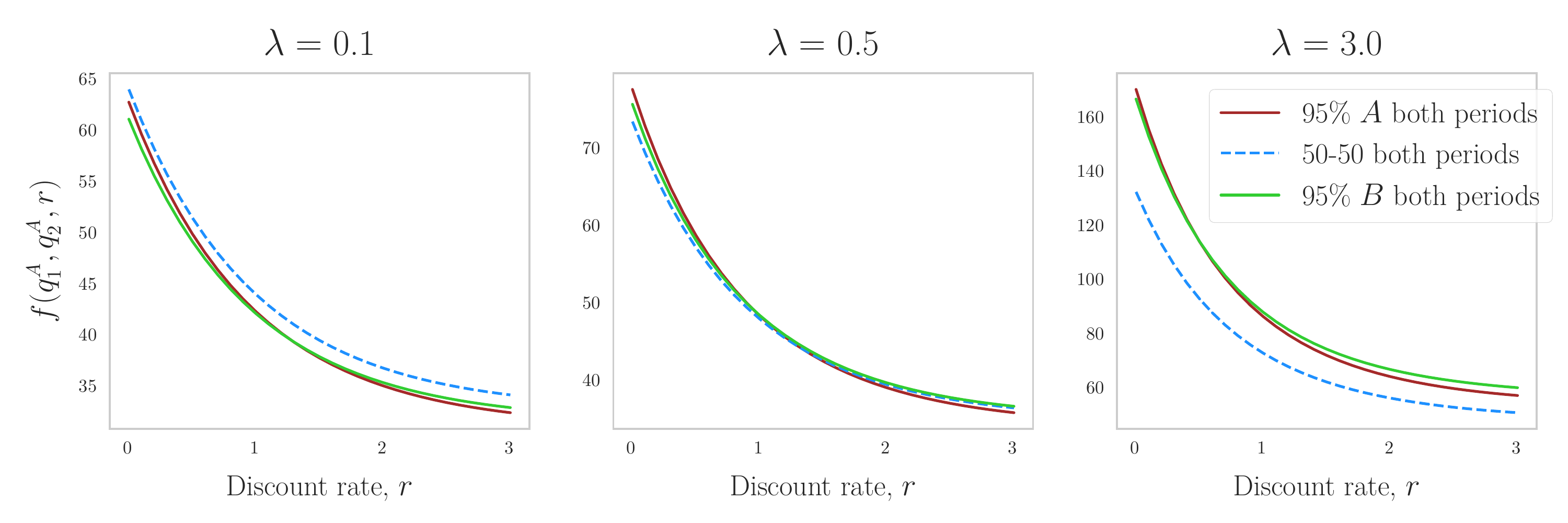}
  \caption{Plots showing how the objective function varies with the discount rate for each of three different technology scenarios, in low, moderate and high risk aversion regimes.
The technologies are the same as in Fig.\ \ref{two_period_contour_plots}.
For low and moderate risk aversion there is a critical discount rate separating regimes in which different scenarios are preferred.}
  \label{two_period_discount_rate}
\end{figure}

Suppose that due to extra system constraints, only three scenarios are available: 95\% technology $A$ in both periods, 50\% technology $A$ in both periods and 95\% technology $B$ in both periods.
Fig.\ \ref{two_period_discount_rate} shows three different risk aversion regimes ($\lambda = 0.1, 0.5, 3.0$), and in each regime the objective function is shown as a function of discount rate, for each scenario.
This shows how the discount rate affects the preference ordering of the three available scenarios in each risk aversion regime.
These plots are readily related back to Fig.\ \ref{two_period_contour_plots} above.

When risk aversion is low the preferred scenario is 95\% technology $B$ in both periods for mild discounting, but as discounting increases this becomes less advantageous, and there is a critical discount rate above which the preferred scenario becomes 95\% technology $A$ in both periods.
When risk aversion is moderate, 50\% $A$ in both periods is preferred for low discounting, all three scenarios are roughly equal when discounting is moderate, and 95\% $A$ in both periods is preferred when discounting is high.
When risk aversion is high a 50-50 mix of technologies in both periods is the preferred scenario regardless of discount rate.

The same arguments and explanations given in Section \ref{section:2period_discussion} apply here.
Clearly this is only a stylized model, with limited scenarios, but the key point is that in some circumstances a critical discount rate exists separating regimes in which different scenarios are preferred.
While this type of result is well known in deterministic situations, and seems trivial in a low dimensional stochastic model such as this, in a multi-technology, multi-period setting this technique could provide essential insight into preference orderings of different technology scenarios.

\section{Conclusion}
\label{section:conclusion}

In this paper we considered two technologies following stochastic experience curves and we characterised the optimal investment (or production) strategies. We used an objective function that accounts for total portfolio cost and uncertainty. The optimal investment depends on risk-aversion, initial conditions (relative technology maturity and initial cost), progress characteristics (mean progress rate and uncertainty of future shocks) and market size; and in the multi-period case also the discount rate.

In contrast to classical Markowitz portfolios, in our setting investment lowers marginal cost, creating a larger and larger incentive to continue investing in the same option. But contrary to the deterministic case, in which these self-reinforcing effects lead to complete specialization, we find that accounting for uncertainty and risk aversion promotes diversification even if one option has better intrinsic technological characteristics. Our results therefore show how the choice of specializing or diversifying depends on the underlying parameters and initial conditions.
Crucially, we find that the nonlinearity of the problem leads to multiple local optima, so that very different optimal portfolios can exist simultaneously, and the global optimum switches instantaneously between them as parameters change. This means that inside a critical region of parameter space a small change in one of the parameters can lead to a very significant change in the optimal portfolio. 

We established an analytical connection between portfolios of technologies and portfolios of financial assets. The Wright's law model of endogenous technological change may be expanded in a series approximation, the leading terms of which are equivalent to those found in the Markowitz model for financial assets. Only the higher order nonlinear terms contribute towards the learning feedback, and hence the Markowitz model may be regarded as the no-learning limit of the technology portfolio model. We also showed that the strength of learning feedback in the model depends on the complete set of model parameters, but in particular on technologies' previous cumulative production and the level of total future demand.
As a result, each technology may be viewed as existing somewhere on a spectrum between more asset-like and more technology-like, depending on the specific set of parameters present.

We also considered the two-period case and found that in a scenario comparison setting (i.e.\ when comparing a limited number of available portfolios, as would be typical in higher-dimensional applications), the discount rate plays a critical role.
For a given level of risk aversion, the discount rate determines which of the available portfolios is preferred.

These findings help establish a theoretical basis for understanding how technologies behave in a simple mean-variance framework, and give insight into how multiple optima can arise in the context of uncertain endogenous technological change.

\section*{Acknowledgements}
This project was primarily supported by Partners for a New Economy.
The project also received funding from the European Union's Horizon 2020 research and innovation programme under grant agreement No.\ 730427 (COP21 RIPPLES), and from the European Research Council under the European Union's Seventh Framework Programme (FP7/2007-2013)/ ERC grant agreement No.\ 611272 (GROWTHCOM).
The authors gratefully acknowledge all these sources of financial support, and additionally the Institute for New Economic Thinking at the Oxford Martin School for its continuing support.
We would also like to thank Cameron Hepburn, Alex Teytelboym, Torsten Heinrich, Jessika E.\ Trancik and B\'ela Nagy for helpful comments and discussions.

\vspace{1cm}
\section*{Appendix}
\appendix

\section{Series expansion approximation to $f$}
\label{appendix:expansion}
The standard Maclaurin series expansion for $(1+x)^{\alpha}$ is
\begin{eqnarray}
(1+x)^{\alpha} &=& \sum_{n=0}^{\infty} \binom{\alpha}{n} x^n \quad \text{for } |x|<1 \label{series_expansion} \\
&=&  1 + \alpha x + \frac{\alpha (\alpha-1)}{2!} x^2 + \frac{\alpha (\alpha-1)(\alpha-2)}{3!} x^3 + \ldots
\end{eqnarray}
where $\binom{\alpha}{n}$ are the generalized binomial coefficients
\begin{equation}
\binom{\alpha}{n} = \frac{\alpha (\alpha-1) \ldots (\alpha-n+1)}{n!}, \quad \alpha \in \mathbb{C}.
\end{equation}
Setting $x$ to $q^i/z_0^i$ (with $q^i< z_0^i$) and $\alpha$ to $-\alpha^i$ gives
\begin{equation}
\left(1 + \frac{q^i}{z_0^i} \right)^{-\alpha^i} = 1 - \alpha^i \left(\frac{q^i}{z_0^i}\right) + \frac{\alpha (\alpha+1)}{2} \left( \frac{q^{i}}{z_0^i} \right)^2 - \frac{\alpha (\alpha+1)(\alpha+2)}{6}\left( \frac{q^i}{z_0^i} \right)^3 + \ldots \ .
\label{simple_expansion}
\end{equation}
Then from Eq.\ (\ref{expandable_f}) we have
\begin{equation}
f =  \sum_{i=A,B} c_0^i q^i \left(1 + \frac{q^i}{z_0^i} \right)^{-\alpha^i} e^{ (\sigma^i)^2 / 2}  + \lambda (c_0^i q^i)^2 \left(1 + \frac{q^i}{z_0^i} \right)^{-2\alpha^i} e^{ (\sigma^i)^2} \left( e^{(\sigma^i)^2} -1 \right), \nonumber
\label{expandable_f2}
\end{equation}
and hence
\begin{eqnarray}
f &=&  \sum_{i=A,B} c_0^i q^i \left( 1 - \alpha^i \left(\frac{q^i}{z_0^i}\right) + \frac{\alpha (\alpha+1)}{2} \left( \frac{q^{i}}{z_0^i} \right)^2 - \dots \right) e^{ (\sigma^i)^2 / 2} \label{expanded_f} \\
&&  \qquad \qquad + \quad \lambda (c_0^i q^i)^2 \left( 1 - 2 \alpha^i \left(\frac{q^i}{z_0^i}\right) + \dots \right) e^{ (\sigma^i)^2} \left( e^{(\sigma^i)^2} -1 \right). \nonumber
\end{eqnarray}
Thus if $q^i$ is much smaller than $z_0^i$, then the approximation formed by discarding everything except the zero order terms in the series expansions will be reasonable. This gives the lowest order approximation to $f$,
\begin{equation}
f \approx \sum_{i=A,B} c_0^i q^i e^{ (\sigma^i)^2 / 2} + \lambda \left( c_0^i q^i \right)^2 e^{ (\sigma^i)^2} \left( e^{(\sigma^i)^2} -1 \right),
\label{appendix_markowitzapprox}
\end{equation}
which has the same form as the Markowitz objective function, with $c_0^i e^{ (\sigma^i)^2 / 2}$ playing the role of expected return and $( c_0^i )^2 e^{ (\sigma^i)^2} \left( e^{(\sigma^i)^2} -1 \right)$ the role of the variance. Note that although the series expansion Eq.\ (\ref{simple_expansion}) converges provided $q^i < z_0^i$, and is itself reasonably approximated by its zero order term (i.e.\ 1) provided $q^i \ll z_0^i$, the lowest order approximation to $f$ (Eq.\ (\ref{appendix_markowitzapprox})) involves discarding the $\frac{(q^i)^2}{z_0^i}$ term in the expectation part of Eq.\ (\ref{expanded_f}), so is reasonable only when $(q^i)^2 \ll z_0^i$. Thus the correct condition for the full optimization problem to be Markowitz-like is $K^2 \ll z_0^i \ \forall i$. Higher order approximations to $f$, and the corresponding conditions for validity, may be computed similarly. See e.g.\ \citet{kraus1976} for further details on portfolio selection involving preferences over skewness.

\section{Analytical points to accompany Section \ref{section:analytics}}
\label{appendix:technical_points}

\subsection{Onset of diversification}
\label{subsection:onset_of_diversification}
For a given pair of technologies the risk neutral portfolio ($\lambda=0$) always concentrates production entirely in one technology due to increasing returns, while for sufficiently large risk aversion the portfolio is diversified over both technologies. The value of risk aversion at which the transition between these two regimes occurs (i.e.\ the onset of diversification) may be found analytically by calculating the intersection of corner and interior solutions. This is done by substituting the relevant boundary condition ($q^A=0$ or $q^A=K$) in the first-order condition equation ($f'(q^A)=0$), and solving for $\lambda$.

For example, consider what happens when $\alpha^B \approx 0.8$ in Fig.\ \ref{alpha2-lambda-surface}. 
For low risk aversion the optimal portfolio is $100\%$ technology $B$, but as $\lambda$ increases there is a value, $\lambda_{\textit{diversification}}$, at which technology $A$ first enters the portfolio. Setting $f'(0)=0$ (with $f$ given by Eq.\ (\ref{obj_fn_explicit})) and solving for $\lambda$ gives
\begin{eqnarray}
\lambda_{\textit{diversification}} (\alpha^B) & = & \frac{c_0^A e^{(\sigma^A)^2 /2}  -  c_0^B \left( \frac{z_0^B}{z_0^B + K} \right)^{\alpha^B} e^{(\sigma^B)^2 /2} \left( 1- \frac{\alpha^B K}{z_0^B + K}  \right)}{2 K (c_0^B)^2 \left( \frac{z_0^B}{z_0^B + K} \right)^{2 \alpha^B} e^{(\sigma^B)^2} \left(e^{(\sigma^B)^2} - 1 \right) \left( 1- \frac{\alpha^B K}{z_0^B + K}  \right)} \\
& = & \frac{\mathbb{E} \left[ c_1^A (0) \right] - \mathbb{E} \left[ c_1^B (K) \right] \left( 1- \frac{\alpha^B K}{z_0^B + K}  \right)}
{2 K \text{Var} \left( c_1^B(K) \right) \left( 1- \frac{\alpha^B K}{z_0^B + K}  \right) }.
\end{eqnarray}
Inserting the parameter values from Table \ref{alpha2-lambda-surface-param-values} in this expression and setting $\alpha^B=0.8$ yields the value $\lambda_{\textit{diversification}} = 0.255$, which is of course consistent with Fig.\ \ref{alpha2-lambda-surface}.
Note that this expression is independent of $\alpha^A$, and only has the interpretation given (onset of diversification) in the specific region of parameter space stated --- it does not in itself determine whether $q^A=0$ is a global minimum in the first place; that must be verified separately.
There is an equivalent expression for the onset of diversification starting from the $100\%$ technology $A$ portfolio, computed in precisely the same way.
Having found analytical expressions for these two major features of the surface, this just leaves the surface discontinuity itself, which we consider next.

\subsection{Location of the discontinuity}
\label{subsubsection:discontinuity}

Observe that in Fig.\ \ref{obj-fn-plot1} the value of the global minimum varies continuously with $\alpha^B$. This follows from the smoothness of the objective function. In fact the global minimum varies continuously with all the underlying parameters, in particular over the entire $\alpha^B$-$\lambda$ grid of Fig.\ \ref{alpha2-lambda-surface}. Hence the value that $f$ attains \textit{at} the surface discontinuity in Fig.\ \ref{alpha2-lambda-surface} is the same when approached from both sides, so we have $f(q_*^A) = f(K-q_*^A)$. This means that near the discontinuity, portfolios on either side of it are symmetric about a 50\% production share. For example, in Figs.\ \ref{obj-fn-plot1} and \ref{cusp_slice_alphaB} the neighbouring portfolios either side of the discontinuity (i.e.\ the point at which the optimum switches) are seen to be symmetric about 50\%, at approximately 20\% and 80\% shares. Additionally, in the case where $q_*^A$ is known exactly (i.e.\ in the small-$\lambda$, full specialization region, where $q_*^A=0 \text{ or } K$), this insight allows the location of the discontinuity to be calculated analytically.

For example, consider what happens for fixed $\alpha^B = 0.65$.
As risk aversion increases from zero the optimal portfolio switches from $0$ to $100\%$ technology $A$.
The critical value at which this switch occurs, $\lambda_{\textit{switch}}$, is found by solving $f(0)=f(K)$ for $\lambda$, which gives

\begin{eqnarray}
\lambda_{\textit{switch}} & = & \frac{c_0^A \left( \frac{z_0^A}{z_0^A + K} \right)^{\alpha^A} e^{ (\sigma^A)^2 / 2} - c_0^B \left( \frac{z_0^B}{z_0^B + K} \right)^{\alpha^B} e^{ (\sigma^B)^2 / 2}}
{K \left[ (c_0^B)^2 \left( \frac{z_0^B}{z_0^B + K} \right)^{2\alpha^B} e^{ (\sigma^B)^2} (e^{ (\sigma^B)^2} -1) - (c_0^A)^2 \left( \frac{z_0^A}{z_0^A + K} \right)^{2\alpha^A} e^{ (\sigma^A)^2} (e^{ (\sigma^A)^2} -1)  \right]} \nonumber \\
& = & \frac{\mathbb{E} \left[ c_1^A (K) \right] - \mathbb{E} \left[ c_1^B (K) \right]}
{K \left[ \text{Var}\left( c_1^B (K) \right) - \text{Var}\left( c_1^A (K) \right) \right] }.
\end{eqnarray}
Again though, note that this expression only has the interpretation given (location of the discontinuity) in the specific region of parameter space stated --- it does not in itself determine whether $q^A=0, K$ are global minima; this must be verified separately.

The same technique may also be used along the $\alpha^B$ axis, to compute the value $\alpha^B_{\textit{switch}}$ at which the discontinuity occurs, for fixed $\lambda$.
There is no elementary closed-form solution for this in general, though there is for the special case $\lambda=0$: solving $f(0)=f(K)$ for $\alpha^B$, with $\lambda=0$, gives
\begin{equation}
\alpha^B_{\textit{switch}} |_{\lambda=0} = \frac{ \log \left( \frac{c_0^A}{c_0^B} \right) + \frac{1}{2} \left[ (\sigma^A)^2 - (\sigma^B)^2 \right] + \alpha^A \log \left( \frac{z_0^A}{z_0^A + K} \right) }
{\log \left( \frac{z_0^B}{z_0^B + K} \right)}.
\end{equation}
Inserting the relevant parameters we find $\alpha^B_{\textit{switch}}|_{\lambda=0}=0.596$ in Fig.\ \ref{alpha2-lambda-surface}.
For other values of $\lambda$ the solution may be computed numerically. For example, fixing $\lambda=0.1$ and solving $f(0) = f(K)$ for $\alpha^B$ yields the value $\alpha^B_{\textit{switch}}=0.681$, which again matches Fig.\ \ref{alpha2-lambda-surface}.

\section{Demand-risk surface for similar technologies}
\label{appendix:demand_risk_surface_similar_techs}
\begin{figure}
  \centering
  \includegraphics[width=1.0\textwidth]{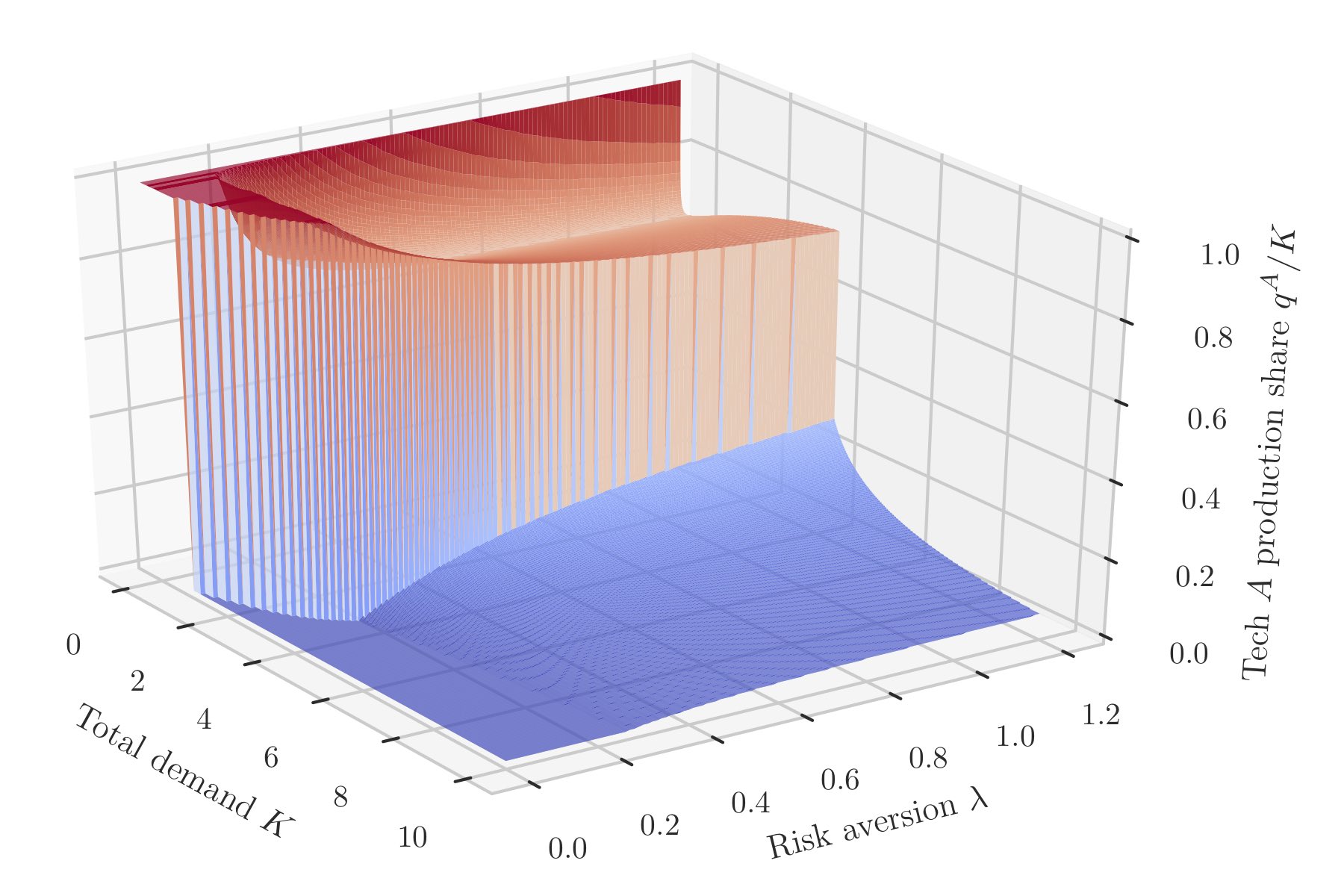}
  \caption{Surface of optimal share of technology $A$ production as total demand and risk aversion vary, for the two almost identical technologies of Section \ref{section:t1}, with $\alpha^B=0.65$.}
  \label{demand_risk_surface_similar_techs}
\end{figure}

In order to provide some comparison with the case of the two asymmetrical technologies (Fig.\ \ref{lambda_K_surface}), Fig.\ \ref{demand_risk_surface_similar_techs} shows the analogous surface for the two almost identical technologies of Section \ref{section:t1}, with $\alpha^B=0.65$ (so technology $B$ has slightly higher experience exponent but also slightly higher noise variance).
Since the technologies have such similar characteristics, there are very few areas of the $K$-$\lambda$ space shown in which complete specialisation is optimal. (Although the $K$ axis scale is different, the other features roughly match Fig.\ \ref{lambda_K_surface}, so the comparison is justified.) In contrast to Fig.\ \ref{lambda_K_surface}, the optimal mix is now diversified (except for very small $K$) for any value of $\lambda$ above around $0.2$, and either technology can dominate, depending on $K$.
Fig.\ \ref{cusp_slice_K} is precisely the $\lambda=0.25$ section through this plot, which explains the nearly-Markowitz origin of the ``ledge'' observed here for very small $K$.

\section{Two-period objective function}
\label{appendix:2period_obj_fn}
Here we provide expressions for the expectation and variance components of the two-period objective function.
We have
\begin{equation}
\label{obj_fn_simplest_form3}
f(q_1^A, q_2^A) = \mathbb{E} \left[ V(q_1^A, q_2^A) \right] + \lambda \text{Var} \left( V(q_1^A, q_2^A) \right),
\end{equation}
where
\begin{eqnarray}
V(q_1^A, q_2^A) & = & \sum_{t=1,2} \sum_{i=A,B} e^{-r(t-1)} c_t^i q_t^i \\
&  = & \sum_{i=A,B} c_1^i q_1^i + e^{-r} c_2^i q_2^i
\end{eqnarray}
is the total system present discounted cost, and first and second period technology costs are given by
\begin{equation}
c_1^i = c_0^i \left( \frac{z_0}{z_1} \right)^{\alpha^i} e^{\eta_1^i} \quad \text{and} \quad c_2^i  \ = \ c_0^i \left( \frac{z_0}{z_2} \right)^{\alpha^i} e^{\eta_1^i + \eta_2^i}
\label{cost3}
\end{equation}
respectively. The noise shocks, $\eta_1^A, \eta_2^A \sim \mathcal{N}(0, (\sigma^A)^2)$ and $\eta_1^B, \eta_2^B \sim \mathcal{N}(0, (\sigma^B)^2)$, are all independent.
Now, in the one period problem independence of technologies means there is no covariance term in $\text{Var} \left( V \right)$, but here, even though technologies are still independent, the second period cost of each technology depends on its own first period cost, so there are nonzero covariance terms to consider.
To simplify notation write
\begin{equation}
\bar{c}_t^i = c_0^i \left( \frac{z_0}{z_t} \right)^{\alpha^i}
\end{equation}
so that
\begin{equation}
c_1^i = \bar{c}_1^i e^{\eta_1^i} \quad \text{and} \quad c_2^i  \ = \ \bar{c}_2^i e^{\eta_1^i + \eta_2^i}.
\label{cost4}
\end{equation}
The following results are then required:
\begin{eqnarray}
\mathbb{E} \left[ e^{\eta_1^i} \right] &=& e^{(\sigma^i)^2/2} \\
\mathbb{E} \left[ e^{\eta_1^i + \eta_2^i} \right] &=& e^{(\sigma^i)^2}\\
\text{Var} \left( e^{\eta_1^i} \right) &=& e^{(\sigma^i)^2} \left( e^{(\sigma^i)^2} -1 \right) \\
\text{Var} \left( e^{\eta_1^i + \eta_2^i} \right) &=& e^{2( \sigma^i)^2} \left( e^{ 2 (\sigma^i)^2} -1 \right) \\
\text{Cov} \left( e^{\eta_1^i}, e^{\eta_1^i + \eta_2^i} \right) &=& e^{3 (\sigma^i)^2/2} \left( e^{(\sigma^i)^2} -1 \right).
\end{eqnarray}
The expectation component of $f$ is thus given by
\begin{eqnarray}
\mathbb{E} \left[ V \right] & = & \mathbb{E} \left[ \sum_{i=A,B} c_1^i q_1^i + e^{-r} c_2^i q_2^i \right] \\
& = & \sum_{i=A,B} \mathbb{E} \left[ c_1^i q_1^i \right] + \mathbb{E} \left[ e^{-r} c_2^i q_2^i \right] \\
& = & \sum_{i=A,B} \mathbb{E} \left[ \bar{c}_1^i q_1^i e^{\eta_1^i} \right] + \mathbb{E} \left[ e^{-r} \bar{c}_2^i q_2^i e^{\eta_1^i + \eta_2^i} \right] \\
& = & \sum_{i=A,B} \bar{c}_1^i q_1^i e^{(\sigma^i)^2/2} + e^{-r} \bar{c}_2^i q_2^i e^{(\sigma^i)^2}, \label{full_two_period_expectation}
\end{eqnarray}
and the variance component by
\begin{eqnarray}
\text{Var} \left( V \right) & = & \text{Var} \left( \sum_{i=A,B} c_1^i q_1^i + e^{-r} c_2^i q_2^i \right) \\
& = & \sum_{i=A,B} \text{Var} \left( c_1^i q_1^i + e^{-r} c_2^i q_2^i \right) \\
& = & \sum_{i=A,B} \text{Var} \left( \bar{c}_1^i q_1^i e^{\eta_1^i} + e^{-r} \bar{c}_2^i q_2^i e^{\eta_1^i + \eta_2^i} \right) \\
& = & \sum_{i=A,B} \left( \bar{c}_1^i q_1^i \right)^2 \text{Var} \left( e^{\eta_1^i} \right) + \left( e^{-r} \bar{c}_2^i q_2^i \right)^2 \text{Var} \left( e^{\eta_1^i + \eta_2^i} \right) \nonumber \\
&  & \qquad \qquad \qquad \qquad \qquad + \ 2 e^{-r} \bar{c}_1^i \bar{c}_2^i q_1^i q_2^i \text{Cov} \left( e^{\eta_1^i}, e^{\eta_1^i + \eta_2^i} \right) \\
& = & \sum_{i=A,B} \left( \bar{c}_1^i q_1^i \right)^2 e^{(\sigma^i)^2} \left( e^{(\sigma^i)^2} -1 \right) + \left( e^{-r} \bar{c}_2^i q_2^i \right)^2 e^{2( \sigma^i)^2} \left( e^{ 2 (\sigma^i)^2} -1 \right) \nonumber \\
&  & \qquad \qquad \qquad \qquad \qquad + \ 2 e^{-r} \bar{c}_1^i \bar{c}_2^i q_1^i q_2^i e^{3 (\sigma^i)^2/2} \left( e^{(\sigma^i)^2} -1 \right). \label{full_two_period_variance}
\end{eqnarray}

\bibliographystyle{agsm}
\bibliography{bib-Portfolio.bib}

\end{document}